\documentclass{aa}
\usepackage[utf8]{inputenc}
\usepackage{amsmath}
\usepackage{amssymb}
\usepackage{graphicx} 
\usepackage{epstopdf}
\usepackage{arydshln}
\usepackage{txfonts}
\usepackage{multicol}
\usepackage{multirow} 
\usepackage{xcolor}
    \title{Comprehensive UV and optical spectral analysis of Cygnus X-1}
    \subtitle{Stellar and wind parameters, abundances, and evolutionary implications} 
    
\author{V. Ramachandran\inst{\ref{inst:ari}} \and A.A.C. Sander\inst{\ref{inst:ari}} \and L.\,M.~Oskinova$^{\ref{inst:UP}}$ \and E.\ C.\ Sch\"osser\inst{\ref{inst:ari}}  \and D.~Pauli$^{\ref{inst:UP},\ref{inst:KL}}$ \and W.-R. Hamann$^{\ref{inst:UP}}$ \and L. Mahy$^{\ref{inst:BG}}$ \and M. Bernini-Peron\inst{\ref{inst:ari}} \and M. Brigitte$^{\ref{inst:CZ}}$ \and B. Kubátová$^{\ref{inst:CZ2}}$}
\institute{
      {Zentrum f{\"u}r Astronomie der Universit{\"a}t Heidelberg, Astronomisches Rechen-Institut, M{\"o}nchhofstr. 12-14, 69120 Heidelberg\label{inst:ari}}\\
              \email{vramachandran@uni-heidelberg.de} 
              \and
    {Institut f{\"u}r Physik und Astronomie, Universit{\"a}t Potsdam, Karl-Liebknecht-Str. 24/25, 14476 Potsdam, Germany\label{inst:UP}}
    \and {Institute of Astronomy, KU Leuven, Celestijnenlaan 200D, 3001 Leuven, Belgium\label{inst:KL}}     
    \and {Royal Observatory of Belgium, Ringlaan 3, 1180 Brussels, Belgium\label{inst:BG}}
     \and {Astronomical Institute of the Czech Academy of Sciences, Bo\v{c}n\'{\i} II 1401/1, 14100 Prague 4, Czech Republic\label{inst:CZ} }
    \and {Astronomical Institute of the Czech Academy of Sciences, Fri\v cova 298, 251 65 Ond\v rejov, Czech Republic\label{inst:CZ2} }
}
\date{February 2023}

\abstract{
    Cygnus X-1  contains the only dynamically confirmed black hole in a persistent high-mass X-ray binary in the Milky Way. Previous studies have suggested that the black hole in  \mbox{Cyg\,X-1} is one of the most massive stellar-mass black holes known in an X-ray binary, despite its high-metallicity environment. While the source has been actively investigated, a comprehensive UV and optical spectral analysis of the donor using modern stellar atmosphere models incorporating stellar winds and X-ray ionization has been lacking.
}{
    We aim to determine the stellar parameters, chemical abundances, and wind parameters of the donor star in \mbox{Cyg\,X-1} along with the mass of the black hole. We also aim to investigate the system's current evolutionary state and its future evolution toward a binary black hole system, exploring its potential as a gravitational wave source.
}{
    We used archival high-resolution UV and optical spectra of \mbox{Cyg\,X-1} taken at multiple orbital phases and X-ray states. We employed state-of-the-art, non-local thermodynamic equilibrium (non-LTE), Potsdam Wolf-Rayet (PoWR) atmosphere models that account for stellar winds, X-ray photoionization, metal line blanketing, and wind clumping. We performed a simultaneous analysis of UV and optical spectra. We further used the stellar evolution code MESA to model the further evolution of the system.
}{
    Our analysis yields notably lower masses for both the donor ($\approx 29\,M_\odot$) and the black hole ($12.7$ to $17.8\,M_\odot$, depending on inclination), and confirms that the donor's radius is close to reaching the inner Lagrangian point. We find super-solar Fe, Si, and Mg abundances (1.3-1.8 times solar) at the surface of the donor star, while the total CNO abundance remains solar despite evidence of CNO processing (N enrichment, O depletion) and He enrichment. This abundance pattern is distinct from the surrounding Cyg OB3 association. We observed a clear difference in wind parameters between X-ray states: $\varv_\infty \approx 1200\,\mathrm{km\,s}^{-1}$ and $\dot{M} \approx 3\times 10^{-7}\,M_\odot\,\mathrm{yr}^{-1}$ in the high-soft state, increasing to $\varv_\infty\lesssim 1800\,\mathrm{km\,s}^{-1}$ and $\dot{M}\lesssim 5\times 10^{-7}\,M_\odot\,\mathrm{yr}^{-1}$ in the low-hard state. The observed X-ray luminosity is consistent with wind-fed accretion. Evolutionary models show that \mbox{Cyg\,X-1} will undergo Roche-lobe overflow in the near future. Under a fully conservative mass accretion scenario, our models predict a future binary black hole merger for \mbox{Cyg\,X-1} within $\sim5$ Gyr. 
}{
    Our comprehensive analysis provides refined stellar and wind parameters of the donor star in \mbox{Cyg\,X-1}, highlighting the importance of using advanced atmospheric models and considering X-ray ionization and wind clumping. The observed abundances suggest a complex formation history involving a high initial metallicity. The potential for a future gravitational wave merger under highly conservative mass accretion makes Cyg X-1 crucial for understanding binary evolution.
} 
  
\begin{document}

\maketitle

\section{Introduction}

Cygnus X-1 is one of the most well-studied high-mass X-ray binaries (HMXBs) and the only dynamically confirmed, persistent black hole (BH) HMXB in the Milky Way. It consists of a BH in a 5.6-day orbit around an O9.7\,Iab supergiant star, HD\,226868 \citep{Walborn1973,Brocksopp1999}.
Recent studies have yielded more precise orbital and distance measurements of the system, suggesting a BH mass of $M_\bullet \approx 21\,M_\odot$, thereby establishing \mbox{Cyg\,X-1} as the most massive BH identified in an X-ray binary \citep[e.g.,][]{Miller-Jones+2021}. 

To derive the BH mass in an X-ray binary, it is crucial to accurately estimate the mass of the optical companion. This requires a detailed multi-wavelength spectroscopic analysis of the donor star using stellar atmosphere models. Although \mbox{Cyg\,X-1} has been extensively studied in X-rays, there are relatively few studies dedicated to the optical and UV spectral analysis of the donor star to derive its fundamental parameters. Most of these studies have employed simplified approaches, such as models containing only hydrogen and helium \citep{Herrero1995}, plane-parallel models  \citep{Caballero-Nieves2009,Shimanskii2012}, and local thermodynamic equilibrium (LTE) models \citep{Orosz2009, Miller-Jones+2021}, without considering any effects of stellar wind. However, massive O stars possess strong stellar winds that significantly affect their spectral features. When simplified approaches are used in spectral analysis, they directly influence the derived stellar parameters, such as surface gravities, and hence directly affect the mass estimation.
Accurate BH mass estimates in \mbox{Cyg\,X-1} are essential for constraining massive binary evolution and understanding the formation of massive BHs at solar metallicity.  In this work, we present a detailed spectral analysis of the donor star in Cyg X-1, considering stellar wind, to derive more accurate donor and BH mass estimates.
 
\mbox{Cyg\,X-1} is a prolific source of X-rays, which are generated as the BH accretes mass from the stellar wind of its companion star. This accretion process results in complex interactions between the stellar wind and the BH’s gravitational field, leading to various observable phenomena across the electromagnetic spectrum. Our current understanding of the wind of massive stars in BH binaries is mostly based on the \mbox{Cyg\,X-1} system. However, the UV spectra of this prototypical system has not been analyzed with the current generation of expanding stellar atmosphere codes, and previous studies are limited to the Sobolev with exact integration (SEI) method \citep{Gies2008,Vrtilek2008}.  Wind mass loss is a key factor in stellar evolution, and hence, a precise measurement of the donor star's mass-loss rate is essential for understanding the evolutionary pathways of these systems. The presence of wind clumping adds complexity to these measurements. Furthermore, in HMXBs, the mass-loss rate and wind structure of the donor star significantly influences the accretion flow onto the compact object. \mbox{Cyg\,X-1} exhibits distinct X-ray states, transitioning between the "low-hard" and "high-soft" states \citep{Liang1984SSRv}, which are defined by their hard and soft X-ray spectra, respectively. While the trigger for these transitions is still debated, changes in the wind accretion rate are a potential contributing factor \citep[e.g.,][]{Gies2003wind}. \mbox{Cyg\,X-1}, therefore, provides an important testbed for investigating these processes. Recent work on the similar HMXB system \mbox{M33\,X-7}, using phase-resolved combined UV and optical spectroscopy coupled with sophisticated atmosphere models, has demonstrated the potential for significant revisions in derived system parameters and wind properties \citep{ramachandran2022}. This highlights the potential for new insights and advancements in our understanding of \mbox{Cyg\,X-1} through a similarly detailed analysis.

Therefore, in this paper, we present a combined UV and optical spectroscopic analysis using detailed atmosphere models to address the current gap in our understanding of \mbox{Cyg\,X-1}. The distance measurement, orbital parameters, and X-ray luminosities of the system are adopted from the literature as listed in Table\,\ref{table:literature}. 
The archival multiwavelength spectroscopic observations are described in Sect.\,2. The details of the atmosphere modeling are given in Sect\,3. The results of the spectral analysis, including fundamental properties, abundances, mass of the components, and stellar wind parameters, are presented in Sect.\,4. Evolutionary implications of the system are discussed in Sect.\,5, and a summary of this study is provided in Sect.\,6.

\begin{table}
\caption{Cyg\,X-1 parameters from the literature adopted in this work.}
    \label{table:literature}
    \centering
    \renewcommand{\arraystretch}{1.4}
\begin{tabular}{lcc}
\hline 
\hline
\vspace{0.1cm}
Parameter                  & Value & Ref. \\
\hline
%RA (J2000)           &  01:33:34.13       \\
%$DEC (J2000)          &  +30:32:11.3       \\
$d$ [kpc]               &  2.22 $\pm$ 0.18  & 1   \\ 
$T_{\mathrm{0}}$ [HJD]                & $2441874.707\pm0.009$     &  2\\
$P_{\mathrm{orb}}$ [days]              &  $5.599829 \pm 0.000016$ & 2 \\ 
eccentricity $e$     &    $0.018\pm0.003$   & 1,3\\
inclination $i$ [$\degr$]      &   $27.5\pm0.7$   &1,3 \\
$f(M)$ [$M_{\odot}$]            &   0.244  &2 \\ 
$L_{\mathrm{x}}\tablefootmark{*}$  [erg\,s$^{-1}$]  &  $\sim$(0.4-4) $\times 10^{37}$    & 4,5  \\
\hline 
\end{tabular}
% \tablebib{
% (1)~\citet{Miller-Jones+2021}; (2) \cite{Brocksopp1999} (3) \cite{Orosz2011} (4) \cite{Wilms2006} (5) \cite{Sugimoto2017}
% }
\tablefoot{
(1)~\citet{Miller-Jones+2021}; (2) \cite{Brocksopp1999} (3) \cite{Orosz2011} (4) \cite{Wilms2006} (5) \cite{Sugimoto2017}
\tablefoottext{*}{X-ray luminosity obtained across various spectral states.}\\ 
}
\end{table}

\section{Spectroscopy}

Archival high-resolution UV and optical spectra, spanning various orbital phases and X-ray states, are used in this study.  Specifically, observations near phases 0 and 0.5 were selected to sample distinct points in the binary orbit and provide a more detailed picture of the system. Two sets of optical spectra were used in this study. The low-hard state sepctra, was retrieved from the IACOB database\footnote{https://research.iac.es/proyecto/iacob/} and obtained with the high-resolution FIbre-fed Echelle Spectrograph (FIES) on the Nordic Optical Telescope (NOT). We also utilized high-soft state optical spectra obtained with the High-Efficiency and high-Resolution Mercator Echelle Spectrograph (HERMES) on the 1.2 m Flemish Mercator Telescope \citep{Raskin2011}. The reduction of the HERMES spectra is described in \citet{Mahy2022}.

For the UV, we primarily utilize spectra acquired with the Hubble Space Telescope (HST) using the Space Telescope Imaging Spectrograph (STIS). These HST UV observations were taken when \mbox{Cyg\,X-1} was in the high-soft state \citep{Gies2008}.
To gain a preliminary understanding of wind at the low-hard state (for which HST UV data do not exist), we used the low-resolution International Ultraviolet Explorer (IUE) far ultraviolet (FUV) spectra at phase 0. 
The details of the spectra used in this work are tabulated in Table\,\ref{tab:data-source}.
Additionally, we used archival FUV and near ultraviolet (NUV)  IUE spectra to construct the spectral energy distribution (SED).  In addition to the spectra, we used various photometric data (from UV to infrared) to build the SED, including U \citep{Bregman1973}; B, and V magnitudes \citep{Zacharias2013}; the Gaia G magnitude  \citep{GaiaCollaboration2022}, and JHK and WISE magnitudes \citep{Cutri2003yCat,Cutri2012wise}.

\begin{table*}
\caption{Spectroscopic data used in this study.}
\label{tab:data-source} 
\renewcommand{\arraystretch}{1.4} 
\begin{tabular}{ccccccccc}
\hline
\hline
 Telescope  & Instrument & $\lambda$ & R     & Date   & Phase  & X-ray spectral state   &  PI\\
    &    &   (\AA)     &      & (yy/mm/dd) & $\phi$ &  &     \\ 
\hline
NOT& FIES &   3730--6865 &46000 & 2008/11/06 & 0.12 & low-hard & Simon-Diaz, S\\
NOT& FIES &   3730--6865 &46000 & 2008/11/08 & 0.47& low-hard & Simon-Diaz, S\\
Mercator & HERMES&    3900--9000 &85000 & 2011/10/11 & 0.009& high-soft & Mahy, L\\
Mercator & HERMES&    3900--9000 &85000 & 2011/10/14 & 0.56& high-soft & Mahy, L\\
HST & STIS/E140M  &  1150--1730  & 45800 & 2002/06/24 & 0.56 &high-soft & Gies, D  \\
HST & STIS/E140M  &  1150--1730  & 45800 & 2002/06/27 &0.09 &high-soft  & Gies, D  \\
IUE& SWP&1250--1900 & 320 &1978/05/10 &0.04 & low-hard& Dupree, A \\
IUE& SWP&1250--1900 & 320 &1978/11/30 &0.02 & low-hard& Dupree, A \\
\hline
\end{tabular}
\end{table*}

\section{Stellar atmosphere modeling}

Stellar atmosphere models have undergone significant advancements in recent decades. Despite being the subject of numerous studies, the current generation of stellar atmosphere models has never been used to conduct a detailed spectral analysis of \mbox{Cyg\,X-1}. The study conducted by \cite{Herrero1995} utilized atmospheric models consisting solely of hydrogen and helium, without taking into account the influence of metal line blanketing.    \cite{Caballero-Nieves2009} employed plane-parallel models from the TLUSTY grid to study optical and UV spectra. Similarly, \cite{Shimanskii2012} also used hydrostatic, plane-parallel stellar atmosphere models to analyze the optical spectra. However, the plane-parallel approximation is suitable only for stars that do not have winds and where the spectrum is solely formed below the sonic point in a quasi-hydrostatic photosphere. 
Given that the donor in \mbox{Cyg\,X-1} is an O supergiant with substantial stellar wind,  a model atmosphere inherently accounting for the wind is highly recommended, as the deduced stellar parameters could otherwise be wrong due to wind emission filling up the absorption profiles. This can have an impact on the derived stellar parameters, such as $\log\,g$. 
To get the parameters of the donor, \cite{Orosz2009} employed LTE models using the generalized model stellar
atmosphere code PHOENIX without considering any effect of stellar winds. \cite{Miller-Jones+2021} followed the same method used in \cite{Orosz2009}, albeit with a revised distance and optimization of parameters. On the other hand, studies such as \citet{Gies2008} and \citet{Vrtilek2008} which focused on studying the wind of \mbox{Cyg\,X-1}, are limited to the SEI method to calculate UV wind line profiles.

Sometimes, HD 226868 is included in samples of stars analyzed by modern non-LTE models, which account for stellar
winds. \citet{Mahy2022} carried out an optical spectroscopic analysis of a sample of  Galactic SB1 systems, including \mbox{Cyg\,X-1}, using CMFGEN models. HD\,226868 is also included in the sample of the automatized analysis by \citet{deBurgos2024} where optical spectra are compared to a large set of FASTWIND models with fixed abundances.

Therefore, in this work, we address these limitations by performing a combined UV and optical spectral analysis of  \mbox{Cyg\,X-1} at  two key orbital phases ($\phi \approx $0 and 0.5) using the state-of-the-art, non-LTE  Potsdam Wolf–Rayet (PoWR) atmosphere model\footnote{http://www.astro.physik.uni-potsdam.de/PoWR/}. 
The PoWR code assumes a spherically symmetric outflow and accounts for iron-line blanketing, wind inhomogeneities, a consistent stratification in the quasi-hydrostatic part, and irradiation by X-rays, all of which are necessary to measure reliable stellar and wind parameters.  To achieve a consistent solution, the equations of statistical equilibrium and radiative transfer are iteratively solved to yield the population numbers while accounting for energy conservation. The radiative transfer is solved in the comoving frame, which avoids simplifications such as the Sobolev approximation. Once an atmosphere model is converged, the synthetic spectrum is calculated via a formal integration along emerging rays. Details of the PoWR code are described in \cite{Graefener2002},  \cite{Hamann2003}, \cite{Todt2015}, and \cite{Sander2015}.

A PoWR model is specified by the stellar temperature $T_\ast$, the bolometric luminosity $L$, the surface gravity $g_\ast$, the mass-loss rate $ \dot{M} $, the wind terminal velocity $\varv_\infty$, the velocity law, and the chemical abundances.  The stellar radius $R_\ast$ (inner boundary) is defined at a Rosseland continuum optical depth of $\tau_{\mathrm {Ross}} = 20$ from $T_\ast$ and $L$. Following the Stefan-Boltzmann law $L = 4 \pi \sigma_{\mathrm{SB}}\, R_\ast^2\, T_\ast^4 $, the stellar temperature  $T_\ast$ is the effective temperature corresponding to the stellar radius $R_\ast$. The outer boundary in our models is set to $R_{\mathrm{max}} =100\,R_\ast$.

In the main iteration, thermal broadening and turbulence are approximately accounted for by assuming Gaussian line profiles with a Doppler width of 30\,km\,s$ ^{-1} $. For the calculation of the emergent spectrum in the formal integral, the Doppler velocity is split into the depth-dependent thermal velocity and a microturbulence velocity $\xi(r)$.  We assume $\xi(r) = \rm max(\xi_{min},\, 0.3\varv(\emph{r}))$ for the models, where the photospheric microturbulent velocity $\rm \xi_{min}$ is varied during the spectral analysis. Pressure broadening is also taken into account.

The velocity field in PoWR models consists of two parts. In the inner part of the stellar atmosphere, the velocity field is calculated consistently such that the quasi-hydrostatic density stratification is fulfilled \citep{Sander2015}. In the supersonic region, the wind velocity field $\varv(r)$ is prescribed assuming a so-called $\beta$-law $\varv(r) = \varv_\infty \left( 1- R_\ast/r \right)^{\beta}$ \citep{CAK1975}, where $ \varv_\infty$ is the terminal wind velocity.  In this work, we explore different values for $\beta$.

In PoWR models, wind inhomogeneities are accounted for in the microclumping approach that assumes optically thin clumps \citep{Hillier1991,Hamann1998} with a void interclump medium and described by a so-called density contrast $D(r)$. The matter density in the clumps is enhanced by a factor $D = 1/f_\mathrm{V}$, where $f_\mathrm{V}$ is the fraction of volume filled by clumps. In this study, we account for a depth-dependent clumping assuming that clumping begins at the sonic point, increases outward, and reaches a density contrast of $D$ at a radius of $R_{\mathrm{D}}$. We adjusted the values of $D$ and $R_{\mathrm{D}}$ during the analysis. 

The models account for complex atomic data of H, He, C, N, O, Mg, Si, P, S, and the iron-group elements. A list of the included ions as well as the number of
accounted levels and line transitions is given in appendix Table\,\ref{tab:modelatom}.
Initially, chemical abundances are adopted from \cite{Asplund2009} and then adjusted to get a better match to the observed spectra.  The iron group elements (Sc, Ti, V, Cr, Mn, Fe, Co, and Ni), with their numerous levels and line transitions, were treated using a superlevel approach \citep{Graefener2002}. This method combines levels and transitions into superlevels with pre-calculated transition cross-sections, assuming solar abundance ratios relative to iron (see appendix Table\,\ref{tab:Fegroup}).
 
\begin{figure}
    \centering
    \includegraphics[width=1\linewidth]{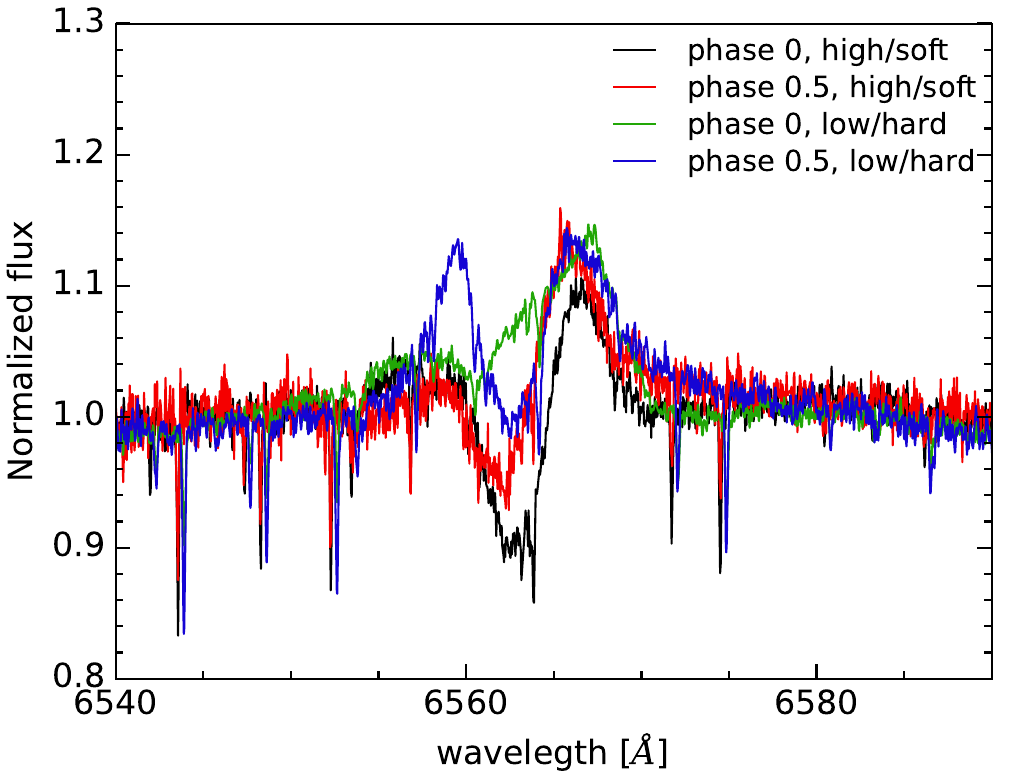}
    \caption{H$\alpha$ line variability observed at different orbital phases and X-ray spectral state of \mbox{Cyg\,X-1} as described in the legend.}
    \label{fig:halpha}
\end{figure}

\begin{figure*}
    \centering 
    \includegraphics[width=0.9\linewidth]{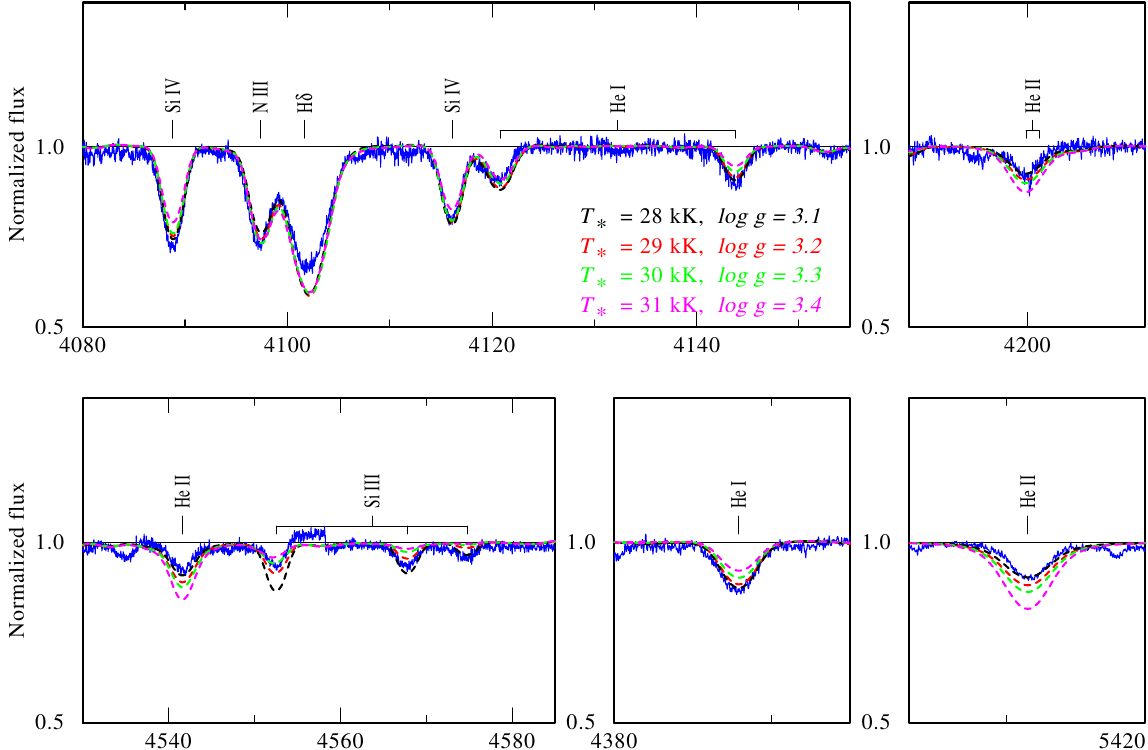}
    \caption{Comparison of observed optical spectra (blue) to synthetic spectra obtained for models with different stellar temperatures and surface gravities (see legend). The plot illustrates the effect of temperature changes on  \ion{Si}{iii} to \ion{Si}{iv} and \ion{He}{i} to \ion{He}{ii} line ratios.}
    \label{fig:teff}
\end{figure*}

\begin{figure}
    \centering
    \includegraphics[width=0.9\linewidth]{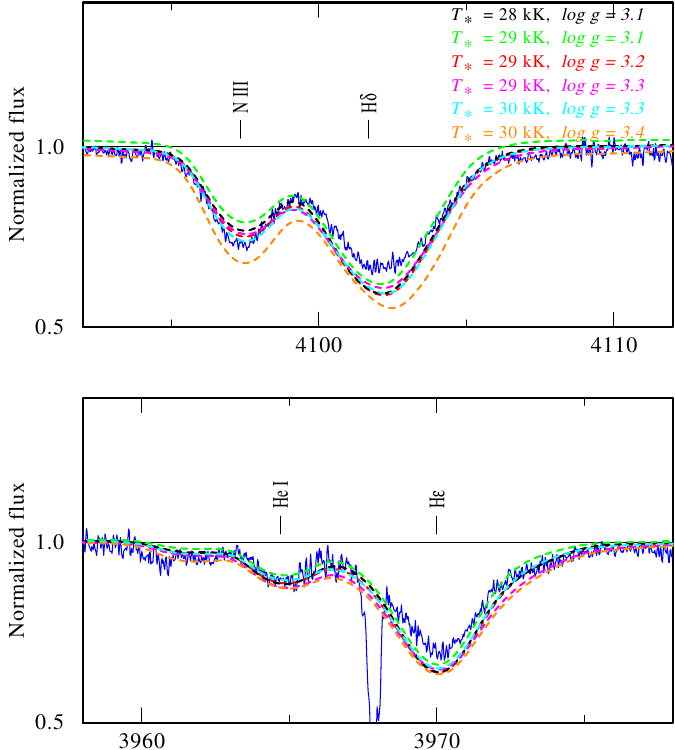}
    \caption{Same as Fig.\,\ref{fig:teff}, but illustrating the impact of $\log g$ on wings of Balmer lines.}
    \label{fig:logg}
\end{figure}

The PoWR code can account for ionization due to X-rays. The X-ray emission is modeled as described 
by \cite{Baum1992}, assuming that the only contribution to the X-ray flux is coming from free--free 
transitions. Since the current generation of PoWR models is limited to spherical symmetry,
the X-rays are assumed to arise from an optically thin spherical shell around the star.
The X-ray emission is specified by three free parameters, which are the fiducial temperature 
of the X-ray emitting plasma $T_{\rm X}$, the onset radius of the X-ray emission $R_0$ $(R_0 > R_\ast)$, 
and a filling factor $X_{\rm fill}$, describing the ratio of shocked to nonshocked plasma.  
We set the onset radius to the orbital distance between the donor star and the BH.
The X-ray temperature and filling factor are adjusted such that the wavelength-integrated X-ray flux from the observations is reproduced by the model. We follow a similar procedure as \cite{Hainich2020}, adjusting input parameters to simulate different X-ray spectral states.

\section{Spectral analysis and results}
\label{sec:spectralanalysis}

We performed the spectral analysis by iteratively fitting the observed spectra with PoWR models. To do that, we begin the analysis with  Milky Way OB-star grid models calculated for solar metallicity with the stellar temperature $T_\ast$ and the surface gravity $\log\,g_\ast$ as main parameters \citep[see][for more details]{Hainich2019}. 
Subsequently, we calculate tailored models with minor adjustments in the stellar parameters and surface abundances to match the observed spectra. This is followed by varying wind parameters in the models to reproduce the UV wind line profiles.   
More details about deriving the stellar and wind parameters as well as chemical abundances, are described in the following subsections. The final parameters derived in this work are given in  Table\,\ref{table:parameters}.

\subsection{Stellar parameters}

The optical spectra obtained at various orbital phases and X-ray spectral states are found to be similar, except for changes in radial velocity and emission components in the Balmer lines, particularly in H$\alpha$. The H$\alpha$ variability is depicted in Fig.\,\ref{fig:halpha}, displaying substantial variations in both the absorption and emission components. Additional emission components from the focused stellar wind can be observed during low-hard spectral states \citep{Yan2008}.  Therefore, we excluded H$\alpha$ from the spectral analysis.  Other Balmer lines also exhibit slight variations across different spectral states, becoming deeper in the high-soft state. On the other hand, all the helium and metal absorption lines and wings of the Balmer lines, are not variable (aside from radial velocity shifts). Hence we focused on these lines to derive stellar parameters and abundances of the donor.

The main diagnostic we use to constrain the temperature of the donor is the He and Si ionization balance based on \ion{He}{i}\,/\,\ion{He}{ii} and \ion{Si}{iii}\,/\,\ion{Si}{iv} line ratios. The following set of optical lines are used for this purpose: \ion{He}{i}~3820, \ion{He}{i}~3866, \ion{He}{i}~4026, \ion{He}{i}~4121,  \ion{He}{i}~4144, \ion{He}{ii}~4200, \ion{He}{i}~4388, \ion{He}{i}~4471, \ion{He}{ii}~4542, \ion{He}{i}~4713, \ion{He}{i}~4920, \ion{He}{i}~5016, \ion{He}{ii}~5412, \ion{He}{i}~5876, \ion{He}{ii}~6670, \ion{He}{i}~7065,  \ion{Si}{iv}~4089, \ion{Si}{iv}~4116, \ion{Si}{iii}~4553, \ion{Si}{iii}~4568, \ion{Si}{iii}~4575, \ion{Si}{iii}~5740.
The pressure-broadened wings of the Balmer lines are the primary diagnostics for surface gravity. We considered H$\epsilon$, H$\delta$, and H$\gamma$ since they are less impacted by wind emissions. Since the ionization balance also reacts to gravity, we simultaneously re-adjusted $T_\ast$ and $\log\,g_\ast$ to achieve a good fit to the observed spectra. A comparison of how models with different temperatures and surface gravities impact the diagnostic line profiles is illustrated in Figs.\,\ref{fig:teff} and \ref{fig:logg}.  The optical spectra are best reproduced by a model with $T_{\ast} = 29$\,kK and $\log g_\ast = 3.2$, which aligns with the spectral classification as a late O-type supergiant.  We noted that our final model better reproduces the depth of Balmer lines at high-soft state (Fig.\,\ref{fig:spectralfitHermes}) compared to low-hard state (Fig.\,\ref{fig:spectralfitIacob}).

Additionally, we varied the micro-turbulence velocity $\xi$ in our models in the range of 10-30\,km\,s$^{-1}$ to fit the overall line profiles of metal lines. The determination of $\xi$ is crucial, so it was constrained along with $T_{\text{eff}}$ and log g during our iterative procedure. An inappropriately chosen microturbulence can lead to substantial shifts in temperature and surface gravity \citep[see, e.g.,][]{Nieva2010}. Consequently, several elements were analyzed to derive $\xi$ and its impact on line profiles of metal lines are shown in appendix Fig.\,\ref{fig:vmic}. We found that the metal lines in the optical and UV spectra are reproduced well with $\xi=22$\,km\,s$^{-1}$. 
With the exception of the automatized solution by \citet{deBurgos2024}, who list $\xi=30$\,km\,s$^{-1}$, this is much higher than in any other previous spectral analysis of \mbox{Cyg\,X-1}: TLUSTY grid models have a fixed $\xi=2$\,km\,s$^{-1}$ for B stars and $\xi=10$\,km\,s$^{-1}$ for O stars; \cite{Orosz2009} and \cite{Miller-Jones+2021} adopted $\xi=2$\,km\,s$^{-1}$ in their grids; \cite{Shimanskii2012} has used a fixed value of $\xi=5$\,km\,s$^{-1}$; and \cite{Mahy2022} used a fixed value of $\xi=10$\,km\,s$^{-1}$ in their CMFGEN grids.  Fixing the micro-turbulence to a lower value will have an impact on the overall spectral fit and the derived stellar parameters and chemical abundances. The fine-tuning of microturbulence is performed along with stellar parameters, and chemical abundances in an iterative manner in this work.
Our derived value is in line with the median value of 20\,km\,s$^{-1}$ obtained by \cite{deBurgos2023} for O9 – B0.5 supergiants in the Galaxy.

For the spectral fit, we adopted the projected rotational velocity ($\varv\sin i$) and macro-turbulent ($\varv_{\mathrm{mac}}$) velocity from \cite{Simon-diaz2017}.
Subsequently, these velocities, along with instrumental broadening, were used to convolve the model spectra to match the observations. Overall, our best-fit model agrees well with the observed UV and optical spectra.

\begin{figure}
    \centering
    \includegraphics[width=0.95\linewidth]{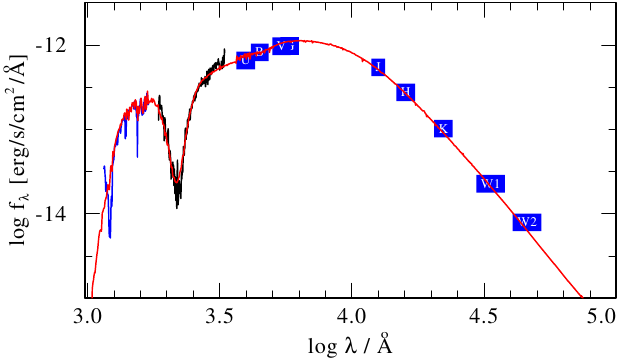}
    \caption{Spectral energy distribution for \mbox{Cyg\,X-1}. The model SED (red solid
line) is adjusted to fit the flux-calibrated HST spectra in the FUV (blue line), IUE spectra in the NUV (black line), and photometry from UV infrared bands (blue boxes).}
    \label{fig:sed}
\end{figure}

The luminosity and color excess $E_\mathrm{B-V}$  were determined by fitting the model SED to the photometry and flux-calibrated UV spectra (see Fig.\,\ref{fig:sed}). We adopted a standard reddening law by \cite{Fitzpatrick1999PASP}.  Subsequently, the model flux is scaled with the adopted distance modulus of 11.73\,mag \citep[based on][]{Miller-Jones+2021}.

\begin{table}
	\caption{Stellar parameters and abundances derived for Cyg\,X-1 in this work. }
	\label{table:parameters}
	\centering
	\renewcommand{\arraystretch}{1.4}
	\begin{tabular}{lc}
		\hline 
		\hline
		\vspace{0.1cm}
		Parameter                                         &          value       \\
		\hline
		\vspace{0.1cm}
		Spectral type                                  & O9.7 Iab                   \\ 
		$T_{\ast}$ (kK)                                & $29^{+1}_{-1}$          \\
		$T_{2/3}$\tablefootmark{1} (kK)                                 & $28.5^{+1}_{-1}$                    \\
		$\log g_\ast$ (cm\,s$^{-2}$)    & $3.2^{+0.1}_{-0.1}$     \\
            $\log g_{2/3}$ (cm\,s$^{-2}$)    & $3.17^{+0.1}_{-0.1}$     \\
		$\log L$ ($L_\odot$)                           & $5.5^{+0.1}_{-0.1}$    \\
		$R_\ast$ ($R_\odot$)                           & $22.3^{+1.5}_{-2.5}$          \\
            $R_{2/3}$ ($R_\odot$)                           & $22.9^{+1.5}_{-2.5}$          \\
		$\varv \sin i$ (km\,s$^{-1}$)       & $112$\tablefootmark{2}       \\
		$\varv_{\mathrm{mac}}$ (km\,s$^{-1}$)       & $81$\tablefootmark{2}       \\
        $\varv_{\rm rot}/\varv_{\rm crit}$         & $\lesssim0.6$       \\
		$\xi$ (km\,s$^{-1}$)            & $22^{+5}_{-5}$          \\
		$X_{\rm H}$ (mass fr.)                         & $0.55^{+0.1}_{-0.1}$    \\
		$X_{\rm He}$ (mass fr.)                         & $0.435^{+0.1}_{-0.1}$    \\
		$X_{\rm C}/10^{-3}$ (mass fr.)                 & $2.2^{+0.5}_{-0.5}$           \\
		$X_{\rm N}/10^{-3}$ (mass fr.)                 & $3.7^{+0.5}_{-0.5}$         \\
		$X_{\rm O}/10^{-3}$ (mass fr.)                 & $3.3^{+0.5}_{-0.5}$           \\
		$X_{\rm Si}/10^{-3}$ (mass fr.)                & $1.2^{+0.5}_{-0.5}$         \\
		$X_{\rm Mg}/10^{-4}$ (mass fr.)                & $9.5^{+3}_{-3}$                      \\
		$X_{\rm Fe}/10^{-3}$ (mass fr.)                & $1.6^{+0.22}_{-0.18}$                     \\
		$E_{B-V}$ (mag)                                & $1.13^{+0.05}_{-0.05}$ \\ 
		$M_\mathrm{donor}$ ($M_\odot$)                  & $29^{+6}_{-3}$        \\
		$M_\mathrm{BH}$ ($M_\odot$)                  & $17.5^{+2}_{-1}$        \\
		$d_{\mathrm{BH}}$ ($R_\ast$)                                 & $2.1^{+0.2}_{-0.1}$     \\ 
		$\varv_\mathrm{orb}$ at BH (km\,s$^{-1}$) & $270^{+30}_{-20}$      \\ 
            $R_\mathrm{eq}\,(R_\odot)$ & $20.2^{+2}_{-1}$\\ 
    $R_{\mathrm{L}1}\,(R_\odot)$ & $26.3^{+2}_{-1}$ \\  
		$\log\,Q_{\mathrm H}$ (s$^{-1}$)   & 48.8   \\
  $\log\,Q_{\mathrm {He\,\textsc{ii}}}$ (s$^{-1}$) & 46--46.5 \\
		\hline
	\end{tabular}
\tablefoot{
Using the derived parameters of the donor and the adopted orbital parameters from literature (see Table\,\ref{table:literature}), we calculated further system parameters. The estimated or assumed mass fractions of elements are also listed. See text for more details.\\
\tablefoottext{1}{$T_{2/3}$ is the effective temperature which refers to the radius where
the Rosseland mean optical depth in the continuum is 2/3}\\
\tablefoottext{2}{taken from \citet{Simon-diaz2017}.}\\
}
\end{table}
\begin{table*}
\caption{Comparison of main stellar parameters of the donor in \mbox{Cyg\,X-1} from literature.} 
    \centering
    \renewcommand{\arraystretch}{1.4}
    \setlength{\tabcolsep}{4pt}
    \begin{tabular}{ccccccc}
    \hline
    \hline
         & This work &Mahy+2022 & Miller-Jones+2021& Shimanskii+2012  & Caballero-Nieves+2009& Herrero+1995\\
         \hline
         $T_{\ast}$ (kK)& 29 &29.8 & 31.14 &30.5 & 28&  32  \\
         $\log g_\ast$ (cm\,s$^{-2}$)& 3.2&3.33 &3.35 &3.31  &3.0 &3.15\\
         $\log L$ ($L_\odot$)&5.5&5.48  & 5.625 & - & 5.45  & 5.4\\
         $R_\ast$ ($R_\odot$)  & 22.3&20.7&22.3& - &22.7 & 17 \\ 
        
         \hline
    \end{tabular}
    
    \label{tab:comparison}
\end{table*}

Our final best-fit model successfully reproduces the observed spectra across both X-ray states and orbital phases. The spectral fits that correspond to high-soft state are in Figures \ref{fig:spectralfitIacob} (phase 0) and \ref{fig:spectralfitIacob05} (phase 0.5), while the same for low-hard state are shown in Figures \ref{fig:spectralfitHermes} (phase 0) and \ref{fig:spectralfitHermes05} (phase 0.5), respectively.  Despite small variations in Balmer line depths, our model provides a good fit to all observations. Table \ref{table:parameters} lists the best-fit stellar parameters and abundances derived in this work for the \mbox{Cyg\,X-1} donor star. Our derived stellar parameters are generally consistent with previous literature estimates (Table \ref{tab:comparison}), though some variations exist.

\subsection{Chemical abundances}

\begin{figure} 
    \includegraphics[width=9cm]{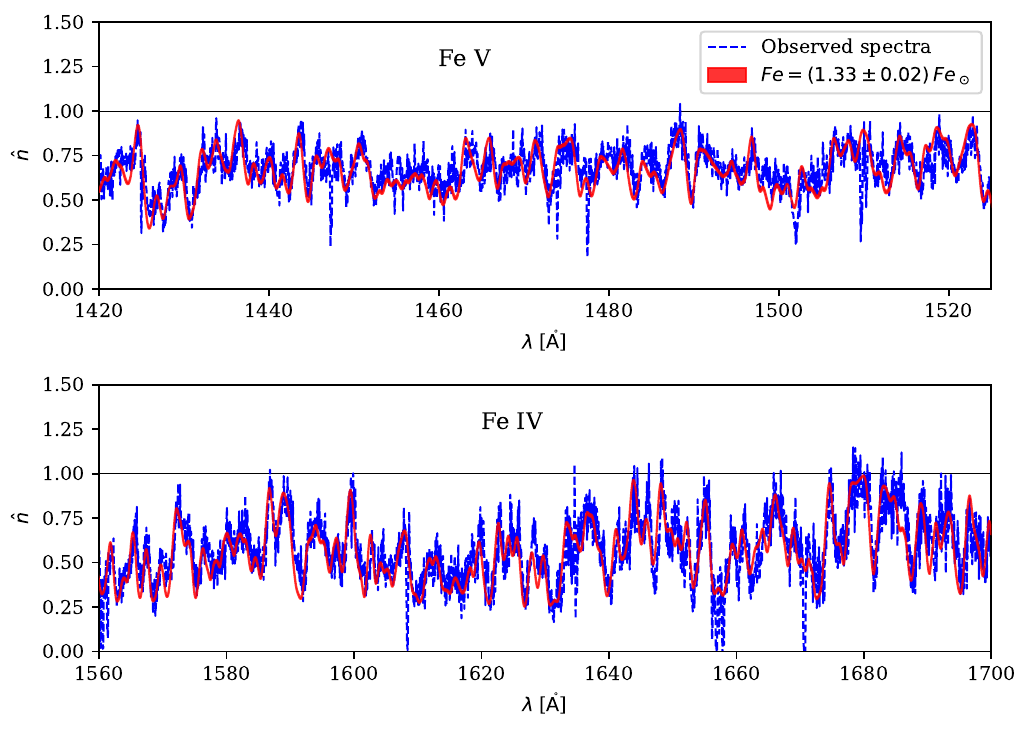}
    \caption{Iron forest in the FUV region. The blue line represents the observed HST spectra, while the best-fit model, obtained using Bayesian analysis, is shown in red for comparison.}
    \label{fig:Fe}
\end{figure}

\begin{table*}
\centering
\caption{Absolute abundances in units of $12 + \log$\,(X$_{i}/$H).}
\label{table:abundance}
\setlength{\tabcolsep}{5pt}
\begin{tabular}{ccccccccccc}
\hline %-------------------------------------------------------------------------------
\hline%-------------------------------------------------------------------------
\noalign{\vspace{1mm}}

& He & C & N  & O & Si & Mg & Fe & $\Sigma$CNO\\
\noalign{\vspace{1mm}}
\hline %-------------------------------------------------------------------------------
\noalign{\vspace{1mm}}
\mbox{Cyg\,X-1} & 11.3  & 8.52 & 8.68 & 8.57 & 7.89 & 7.85 &7.72 & 9.07 \\

\noalign{\vspace{1mm}}
\hdashline
\noalign{\vspace{1mm}}
Sun\tablefootmark{1} & 10.93 & 8.43 & 7.83  &8.69 & 7.51 & 7.6 & 7.5 &8.92\\ 
B stars in solar neighborhood\tablefootmark{2} & 10.99 &  8.33 &7.79 & 8.76 & 7.5 & 7.56 &7.52 &8.93 \\
B stars in Cyg\,OB3\tablefootmark{3} &  & 8.15 & 7.68 & 8.54 & 7.22 & 7.63 &7.33 &8.73 \\ 
\hline%-------------------------------------------------------------------------
\end{tabular}
\tablefoot{
\tablefoottext{1}{from \citet{Asplund2009}}
\tablefoottext{2}{average abundances of  B stars in the  solar neighborhood from \cite{Nieva2012}}
\tablefoottext{3}{average abundances of  B stars in the  Cyg OB3 from \cite{Daflon2001}. Here, Mg and Fe are  LTE abundances, and  C, N, O, and Si are non-LTE abundances.}
}
\end{table*}

\begin{table}[h!]
\centering
\caption{Abundances relative to solar (by mass).  }
\label{table:abundancediff}
\setlength{\tabcolsep}{5pt}
\begin{tabular}{lccccccc}
\hline
\hline
Star & C & N & O & Si & Mg & Fe & $\Sigma$CNO \\
\hline
Cyg X-1 & 0.93 & 5.36 & 0.57 & 1.80 & 1.37 & 1.33 & 1.05 \\
Bstars & 0.79 & 0.91 & 1.17 & 0.98 & 0.93 & 1.12 & 1.05 \\
CygOB3 & 0.52 & 0.71 & 0.71 & 0.51 & 1.10 & 0.72 & 0.66 \\
\hline
\end{tabular}
\end{table}

We adjusted the abundances of individual elements to reproduce the observed strength of their respective optical absorption lines. 
The full list of lines that were considered for the abundance determination are 
\begin{itemize}
    \item Helium:  \ion{He}{i}~3820, \ion{He}{i}~3866, \ion{He}{i}~4026, \ion{He}{i}~4121,  \ion{He}{i}~4144, \ion{He}{ii}~4200, \ion{He}{i}~4388, \ion{He}{i}~4471, \ion{He}{ii}~4542, \ion{He}{i}~4713, \ion{He}{i}~4920, \ion{He}{i}~5016, \ion{He}{ii}~ 5412, \ion{He}{i}~5876, \ion{He}{ii}~6670, \ion{He}{i}~7065.

    \item Carbon: \ion{C}{iii}~1247, \ion{C}{ii}~4267, \ion{C}{iii}~4647-4650, \ion{C}{iii}~5696, \ion{C}{iv}~5801, \ion{C}{iv}~5812.

    \item Nitrogen: \ion{N}{iv}~1483-1487, \ion{N}{ii}~3995, \ion{N}{iii}~4097, \ion{N}{ii}~4237-4242, \ion{N}{iii}~4512-4518, \ion{N}{iii}~4634-4641, \ion{N}{ii}~5667-5676.

    \item Oxygen: \ion{O}{iv}~1339-1343,  \ion{O}{iii}~1409-1412, \ion{O}{iii}~3963, \ion{O}{ii}~4070-4076, \ion{O}{ii}~4185-4190, \ion{O}{ii}~4417, \ion{O}{ii}~4649, \ion{O}{ii}~4676, \ion{O}{iii}~5592.

    \item Silicon: \ion{Si}{iii}~1294-1303, \ion{Si}{iv}~4089, \ion{Si}{iv}~4116, \ion{Si}{iii}~4553, \ion{Si}{iii}~4568, \ion{Si}{iii}~4575, \ion{Si}{iii}~5740

    \item Magnesium: \ion{Mg}{ii}~4481.
\end{itemize}

Individual elemental abundances by mass and by number are given in Tables\,\ref{table:parameters} and \ref{table:abundance} respectively. 
Compared to solar and average abundances of nearby B stars, we found evidence for nitrogen enrichment (by a factor of $\sim 5$) and oxygen deficiency in the \mbox{Cyg\,X-1} donor, while carbon remains the same (see relative abundances in Tables\,\ref{table:abundancediff}). This nitrogen overabundance is consistent with that reported by \citet{Caballero-Nieves2009}. We tested a range of hydrogen mass fractions from $0.4$ to $0.73$ in the study and found that models with $X_\text{H}=0.55$ reproduce the intensity of hydrogen and helium lines in the observed spectra most accurately. This provides evidence for He enrichment at the surface of the \mbox{Cyg\,X-1} donor. \citet{Mahy2022} and \citet{Shimanskii2012} also reported enrichments in He and N in their analysis. 

The surface abundances of massive stars evolve due to mass loss, which exposes deeper layers, and mixing processes, which alter internal compositions. In the case of the \mbox{Cyg\,X-1} donor, both effects likely contributed to the observed He and N enrichment. 
Adopting the inclination of the orbit for the donor's rotation axis, $\varv\sin i$ implies a rotational velocity at the equator $\approx 240$\,km\,s$^{-1}$, suggesting the donor star is rotating at $\approx 0.6\varv_{\rm{crit}}$. Assuming tidal synchronization would suggest a higher inclination of $i\approx 33.7$ (see Sect.\,\ref{sec:mass}) and hence $\varv_{\rm rot}/\varv_{\rm crit} \approx 0.5$.
Such rapid rotation can lead to enhanced mixing. Additionally, interaction with the BH companion may have partially stripped the outer layers of the donor, further contributing to the enrichment.

 Although we found evidence for CNO processed material on the surface of the donor, the overall sum of CNO abundance is in agreement with solar value. The chemical abundances in \mbox{Cyg\,X-1} donor relative to solar are illustrated in Fig.\,\ref{fig:abundace}.  Notably, $\alpha$-elements like Si and Mg are overabundant, by factors of 1.8 and 1.37 respectively \citep[relative to solar abundances from][]{Asplund2009}.  Using more recent solar abundance estimates \citep{Magg2022} these factors slightly change to 1.5 for both Si and Mg.  These $\alpha$-element abundances are also significantly higher than those observed in B stars in the solar neighborhood.

\subsubsection{Fe abundance determination from FUV iron forest}

The forest of photospheric Fe absorption lines in the FUV spectra of hot massive stars provides a direct method to constrain their metal content. Previous studies have assumed a solar-like iron abundance for the system ($X_{\rm Fe}=1.2\times10^{-3}$). In our models, we varied the Fe mass fractions within the range of $1.1 - 2.4 \times10^{-3}$ (iron group mass fraction $X_{\rm G}=1.2-2.6\times10^{-3}$) and employed Bayesian analysis to determine the best-fit values \citep[see][for details on the method]{Schoesser+2025}. We focused on two regions of the FUV spectra, 1420-1520\AA\ and 1560-1700\AA\, which are dominated by \ion{Fe}{v} and \ion{Fe}{iv} lines, respectively.

Our results indicate that the iron forest in \mbox{Cyg\,X-1} UV spectra is best reproduced with models assuming a slightly super-solar Fe abundance of $1.33\,Z_\odot$ ($X_{\rm Fe}=1.6\times10^{-3}$) as illustrated in Fig.\,\ref{fig:Fe}. The derived Fe abundance is higher than the average Fe abundance of B stars in the solar neighborhood \citep{Nieva2012}.  In these models, we used a microturbulence value ($\xi$) of 22 km s$^{-1}$, derived from the optical metal lines.
However, these iron line features are sensitive to the adopted $\xi$ in the model \citep[see, e.g.,][]{Hillier2003}. To determine the possible uncertainty range in the Fe abundance, we varied $\xi$ between 17 and 27 km s$^{-1}$ along with various Fe mass fractions. The final derived Fe abundance is $1.33^{+0.19}_{-0.15}\,Z\odot$. 

Several studies have reported exceptionally high Fe abundances in \mbox{Cyg\,X-1},  $\sim 2.0- 5$ times solar using reflection fits to X-ray spectra \citep{Tomsick2014,Walton2016,Parker2015,Basak2017MNRAS}. Similarly, analysis of the Fe K$\alpha$ emission line indicates Fe enrichment, with estimates of $\sim 1.6 - 6$ times solar \citep{Duro2011,Duro2016}. These results suggest a substantial enrichment of Fe in the accretion disk or the surrounding material.  While these value are significantly higher that our estimated Fe abundance, they are consistent with the general trend of Fe enrichment. The discrepancy may be attributed to the sensitivity to different ionization states of Fe in X-ray and UV, or the underlying physical models used in each analysis.  

\subsubsection{Initial supersolar composition or recent enrichment by supernova}

\begin{figure}
    \centering
    \includegraphics[width=1\linewidth]{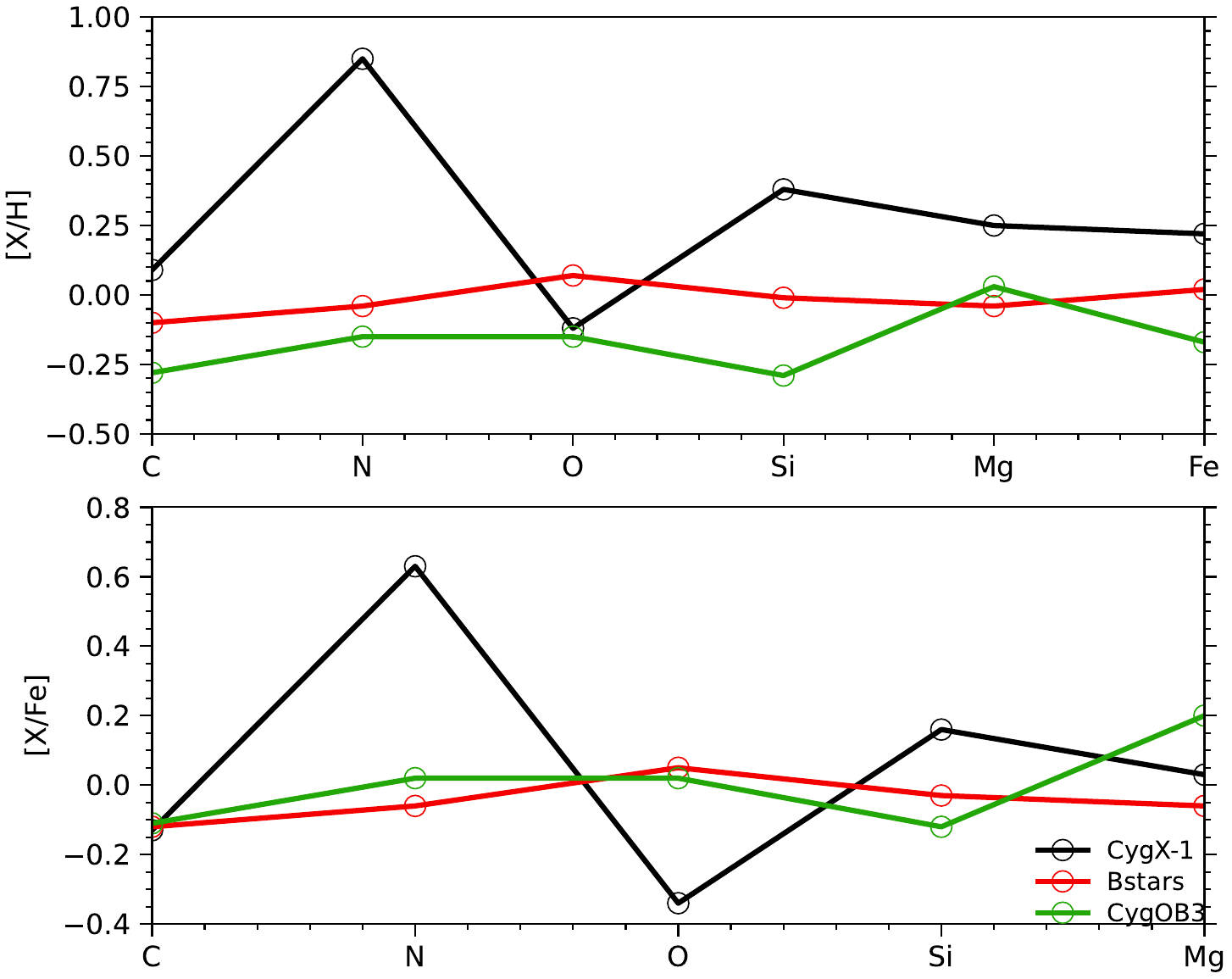}
    \caption{Elemental abundances in \mbox{Cyg\,X-1} donor O star relative to hydrogen (upper panel) and iron (lower panel). The abundance ratios are relative to solar and in logarithmic scale. For comparison, the average chemical abundance of B stars in the Cyg OB3  \citep{Daflon2001} and B stars in the solar neighborhood  \citep{Nieva2012} are marked.}
    \label{fig:abundace}
\end{figure}

In \mbox{Cyg\,X-1}, we found that the Fe, Si, and Mg abundances are approximately 1.3-1.8 times higher (by mass) than the corresponding solar values (0.2-0.3 dex higher by number, see Fig.\,\ref{fig:abundace}). This exceeds the typical metal content of \ion{H}{ii} regions located approximately 7.7 kiloparsecs from the center of the Galaxy. The \mbox{Cyg\,X-1} donor may either have been polluted by material from a supernova explosion, particularly from its companion star, or it may have formed from interstellar medium (ISM) material which was already enriched.
 
\mbox{Cyg\,X-1} is located at the boundary of the Cyg OB3 association. A recent study by \citet{rao2020kinematic} using Gaia astrometric data and radial velocities supports Cyg OB3 as the parent association of \mbox{Cyg\,X-1}, as its kinematic characteristics align with most stars in Cyg OB3. This is also consistent with previous works by \citet{Mirabel2003Sci} and \citet{Mirabel2017IAUS,Mirabel2017NewAR}, indicating that \mbox{Cyg\,X-1} likely formed in situ within the association.

\citet{Daflon2001} measured chemical abundances of slowly rotating B stars in OB associations, including Cyg OB3. The average abundance of B stars in Cyg OB3 is listed in Table\,\ref{table:abundance} for comparison. The Fe and Si abundances they reported are rather lower than solar values, while Mg abundance is comparable. However, the Fe and Mg abundances provided by \citet{Daflon2001} are based on analyses using LTE models and hence have higher uncertainties. A similar under-abundance of elements relative to solar is also reported for stars in the nearby association Cyg\,OB7 by the same study. \citet{Berlanas2018} conducted a study in Cyg OB2 and did not observe any significant rise in [O/H] and [Si/H] levels, leading to the conclusion that there is no evidence of chemical enrichment. While non-LTE Fe determinations of hot stars in the region and in particular of Cyg OB3 are lacking, taking the existing literature at face value makes it puzzling why only the \mbox{Cyg\,X-1} system shows super-solar Fe, Si, and Mg abundances compared to the neighborhood.

Certain studies have identified significant variability in the overall metallicity of stars within the proximate Galactic interstellar medium. For example, \citet{DeCia2021} identified substantial disparities in gas-phase metallicity, with variations reaching up to tenfold, in the neighborhood of the Sun. This was established by examining neutral clouds in the line of sight of 25 O- or B-type stars located in or near the Galactic plane and within 3\,kpc. In contrast, research by \citet{ArellanoCordova2020,ArellanoCordova2021} indicated that the gas in \ion{H}{ii} locations is uniformly distributed at a specific Galactocentric distance, based on chemical abundance studies of B-type stars, classical Cepheids, and young clusters. \citet{Esteban2022} also suggest that the chemical composition of the interstellar medium in the solar vicinity is very homogeneous.

The hypothesis that the BH progenitor polluted the donor star's surface via a supernova explosion has been previously considered. \citet{Shimanskii2012} reported higher Ne and Si abundances relative to solar in the \mbox{Cyg\,X-1} donor, which they suggested as evidence for a supernova event. Our analysis of the surface abundances in \mbox{Cyg\,X-1} reveals enrichment in Fe, Si, and Mg, but not in CNO elements. However, this observed abundance pattern presents a challenge to a pure supernova origin.
While \citet{rodriguez2023iron} estimated a mean Fe yield of core-collapse supernovae (CCSNe) to be $\approx0.058\,M_\odot$, with Type Ic supernovae producing more than twice this amount, this is still significantly lower than the Fe yield of Type Ia supernovae  \citep[$\sim 0.6-0.7\,M_\odot$; e.g.,][]{Howell2009SNIa}. Based on the Fe yields of CCSNe, \citet{weinberg2024scale} inferred corresponding  yields of $\alpha$-elements such as Mg, Si, and O. Considering their estimates, a 33\% increase in Fe abundance would be accompanied by ay a $\sim 30-35$\% increase in Si and Mg, along with an approximately twofold increase in O abundance \citep[see table 1 and figure 3 in][]{weinberg2024scale}.
Even in alternative scenarios like a failed supernova or a direct collapse to a BH, some O enrichment relative to Fe would still be expected. Therefore, the observed abundance pattern in \mbox{Cyg\,X-1} is challenging to explain solely by supernova yields or stellar winds from the BH progenitor.

\subsection{Mass of the components}\label{sec:mass}

\begin{figure}
    \centering
    \includegraphics[width=1\linewidth]{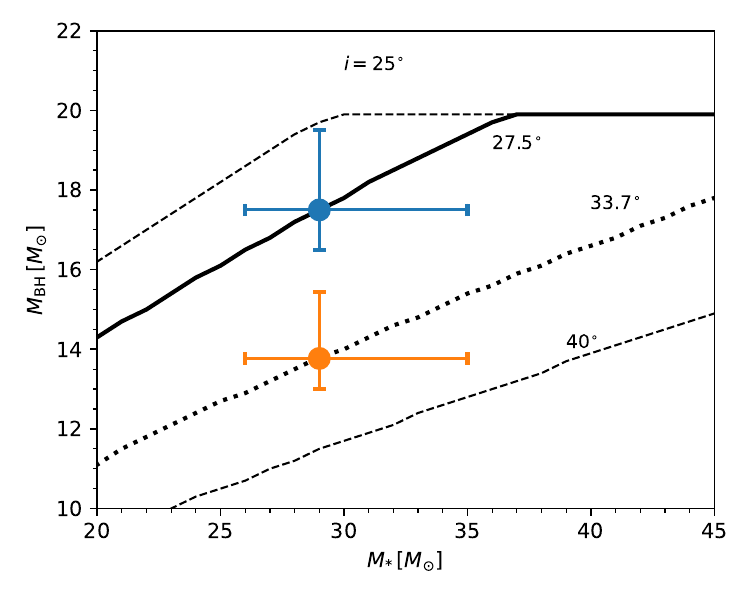}
    \caption{Mass plane diagram for \mbox{Cyg\,X-1}, showing component masses and their uncertainties. BH masses are estimated for $i= 27.5^\circ $(blue) and $i= 33.7^\circ $ (orange).The solid black line represents the mass relation for the adopted inclination of  $i= 27.5^\circ $ derived from optical observations \citep{Orosz2011, Miller-Jones+2021}. The dotted line corresponds to  $i= 33.7^\circ $ derived assuming synchronous rotation. The dashed lines illustrate the variation in the mass relation for alternative inclination angles. }
    \label{fig:mass}
\end{figure}
\begin{figure}
    \centering
    \includegraphics[width=1\linewidth]{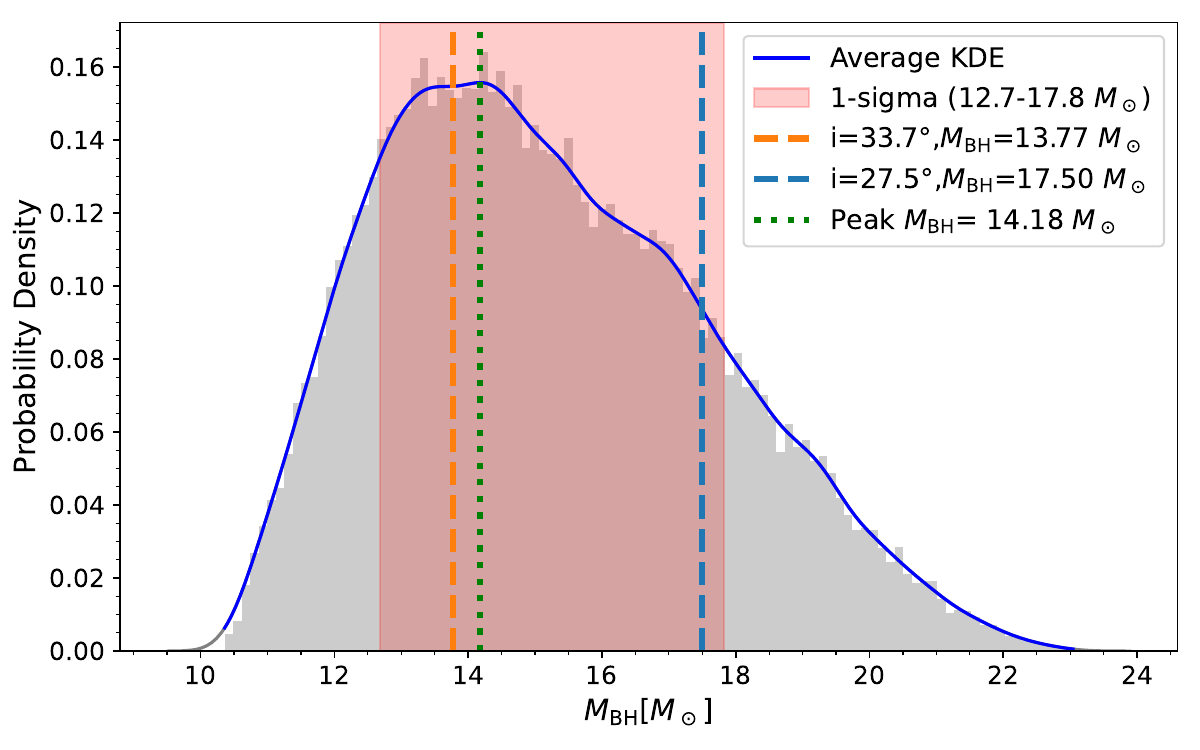}
\caption{Probability distribution of BH mass in \mbox{Cyg\,X-1}. The gray histogram and blue line depict the $M_\mathrm{BH}$ distribution and its KDE, respectively. The shaded region shows the 1-$\sigma$ range (16th and 84th percentiles). The dotted green line marks the KDE peak, and the dashed orange and blue lines represent the $M_\mathrm{BH}$ values at inclinations of $27.5^\circ$ and $33.7^\circ$ (see legend).}
    \label{fig:MBH_distribution} 
\end{figure}

 \begin{figure}
     \centering
     \includegraphics[width=1\linewidth]{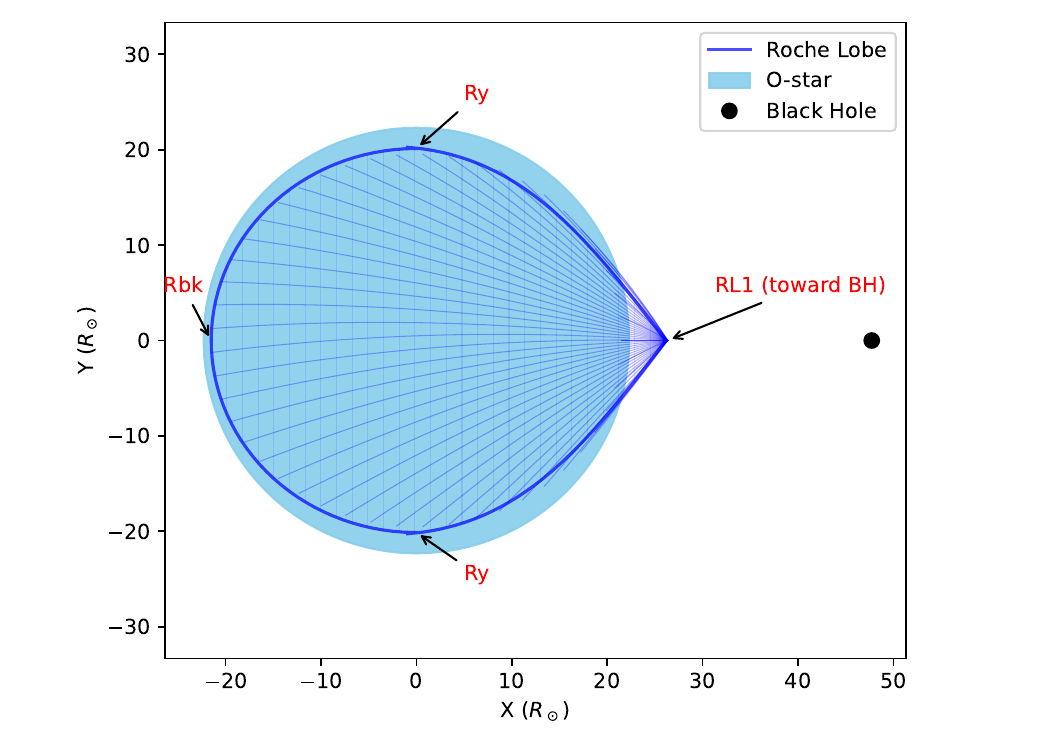}
     \caption{Illustration of the Roche lobe geometry for \mbox{Cyg\,X-1}, showing the O-star donor (light blue circle) and the BH (black dot, for $M_\mathrm{BH} = 17.5\,M_\odot$). The dark blue lines and the mesh depict the approximate Roche equipotential surface, with arrows indicating the distances $R_\mathrm{L1}$, Rbk, and Ry.  The axes are scaled in  $R_\odot$.}
     \label{fig:roche-geo}
 \end{figure}
 The spectroscopic mass of the donor is calculated from $\log\,g_\ast$ and $R_\ast$ ($g_\ast=G\,M_\ast\,R_\ast^{-2}$), and the uncertainties in these parameters propagate into the final uncertainty of the donor mass.
The reduction in luminosity, temperature, and gravity results in a much lower spectroscopic mass of $\approx 29\,M_\odot$ for the O star than previously reported in \cite{Miller-Jones+2021}. Considering the orbital parameters in Table\,\ref{table:literature},  we estimate the BH mass, the orbital radius, and the Roche lobe radius of the donor. The newly derived BH mass is $\approx 17.5\,M_\odot$, which is lower than the previous estimate of $\approx 21\,M_\odot$ by \cite{Miller-Jones+2021}. 

The inclination angle of the binary system \mbox{Cyg\,X-1} is a crucial parameter affecting the BH mass estimate as demonstrated in the  Fig.\,\ref{fig:mass}. While previous optical measurements indicate an orbital inclination of  $i= 27.5^\circ $ \citep{Orosz2011, Miller-Jones+2021}, X-ray reflection and polarization studies suggest a significantly higher disk inclination, ranging from $35^\circ$ to $45^\circ$ \citep[e.g.,][]{Krawczynski2022,zdziarski2024spin,Walton2016,Tomsick2014}. This discrepancy is attributed to a possible disk warp or misalignment between the binary and inner disk axes. If the orbital inclination is higher, then it would lead to significantly reduced BH masses (for instance $M_{\rm{BH}}\approx 11.5\,M_\odot$ at $i\sim 40^\circ$). 

While detailed light curve modeling, which could provide further constraints on the system's geometry and BH mass, is beyond the scope of this paper, we estimate the inclination angle $i$ using the assumption of tidal synchronization. We assume that the primary star's rotational period is synchronized with the binary orbital period, a plausible scenario given the system's short orbital period \citep[see, e.g.,][]{Stoyanov2009AN}. The relationship between $i$, the observed $\varv\sin i$, the primary star radius $R_1$, and the orbital period $P$ is given by
\begin{equation}
  i = \arcsin \left( \frac{\varv \sin i \cdot P}{2 \pi R_1} \right)\text{.}   
 \label{eq:i} 
\end{equation}
Using $\varv\sin i$ of $112$\,km\,s$^{-1}$ (see Table\,\ref{table:parameters}) in Eqn.\,\ref{eq:i} yields an inclination of $i \sim 33.7^\circ$ and BH mass estimate of $M_{\rm{BH}} \approx 13.8\,M_\odot$. The correlation between the estimated masses of the donor star and the BH, along with their respective uncertainties, is illustrated in Fig.\,\ref{fig:mass}.

To further investigate the uncertainty in the BH's mass, we conducted a Monte Carlo simulation by sampling the O-star mass within the 2$\sigma$ range. This wider range of donor masses allowed us to investigate the impact of more extreme $M_{\mathrm{donor}}$ values and to ensure the robustness of our results against potential outliers or underestimated systematic errors.
Given that the uncertainties in donor mass are asymmetric, we utilized a truncated normal distribution to accurately represent its probability distribution. The simulation adopts a uniform probability distribution of inclination in the range $i \sim 25^\circ - 39^\circ$. This range was estimated based on the assumption of synchronous rotation and reflects the spread of $\varv\sin i$ values in the literature ($\varv\sin{i}\sim 90-120$\,km/s).
The resulting $M_\mathrm{BH}$ distribution (adopting a mass function as in Table\,\ref{table:literature}), is shown in Figure \ref{fig:MBH_distribution}. Applying kernel density estimation (KDE) to characterize this distribution reveals a peak at $\approx 14.2\,M_\odot$ with a $1\sigma$ range of masses $12.7$ to $17.8\,M_\odot$. The inclination-driven uncertainty, as shown in the $M_\mathrm{BH}$ distribution, suggests that the $17.5\,M_\odot$ estimate (corresponds to $i= 27.5^\circ$), likely represents an upper limit.  Notably, future astrometry data would provide independent and robust constraints on inclination and mass ratio, significantly reducing uncertainties in BH mass determination.

For $M_{\rm{BH}} \approx 17.5\,M_\odot$, the orbital separation is $47.8\,R_\odot$ while the radius of the donor star is $R_\ast\approx22.3\,R_\odot$. \citet{Eggleton1983} defines $R_\mathrm{eq}$ as an equivalent radius that corresponds to the volume of the Roche lobe when approximated as a sphere.  In this approximation, $R_\mathrm{eq}$ in Cyg X-1 is roughly similar to the donor radius with a Roche lobe filling factor of $f_\mathrm{Req}= R_\ast/R_\mathrm{eq} \approx 1.1^{+0.15}_{-0.2} $. However, these radii are considerably smaller than the distance to the inner Lagrangian point ($R_\mathrm{L1}$).

As further discussed in Sect.\,\ref{sec:windvsroche},  the classical Roche lobe geometry of the donor star is modified in the presence of a radiation-driven stellar wind \citep[see also][]{Dermine+2009}. 
To visualize this scenario, we employed the formalism from \citet{Leahy2015} and constructed a 2D representation of the Roche lobe geometry in Cyg X-1, as shown in Fig.\,\ref{fig:roche-geo}.  We calculated various radii defining the Roche lobe's shape: the distance to the inner Lagrangian point $R_\mathrm{L1}$, the radii along the line connecting the stars (\(R_\text{fr}\), \(R_\text{bk}\)), and the radii perpendicular to this line and the orbital plane (\(R_{y}\), \(R_{z}\)). These radii represent distances to the Roche equipotential surface at specific points, allowing us to map the shape of the Roche lobe. This plot, generated using a mass ratio of $q = M_\text{BH}/M_\text{donor} = 0.603 \approx 17.5/29$, illustrates the relative sizes and positions of the donor star, the Roche lobe, and the position of the BH. This analysis yields $f_\mathrm{R_{L1}} = R_\ast/R_\mathrm{L1} \approx 0.85$. Even with a BH mass of $\approx 13.8\,M_\odot$, this value only changes marginally to $\approx 0.835$. The presumably still rather spherical donor star ($\varv_{\rm rot}/\varv_{\rm crit} \leq 0.6$) extends beyond the $R_\mathrm{eq}$ and the radii in the y, z, and back directions ($R_y$, $R_z$, $R_\text{bk}$), yet remains smaller than the distance to the inner Lagrangian point $R_\mathrm{L1} \equiv R_{fr}$. This holds true across the entire BH mass probability distribution shown in Fig.\,\ref{fig:MBH_distribution} (see the Roche lobe filling factor distribution in various directions in appendix Fig.\,\ref{fig:fRL_hist} ). 
Nonetheless, the donor star is still close to a situation where overflowing mass transfer could happen.
This result is consistent with prior studies \citep{GiesBolton1986, Gies2003wind, Ziolkowski2005evolutionary}.  Similar behavior has been observed in other BH-HMXBs, such as Roche lobe overfilling in M33 X-7  \citep{ramachandran2022} and nearly Roche lobe filling in LMC\,X-1 \citep{Orosz2009}.

\begin{figure*}
    \centering
    \includegraphics[width=1\linewidth]{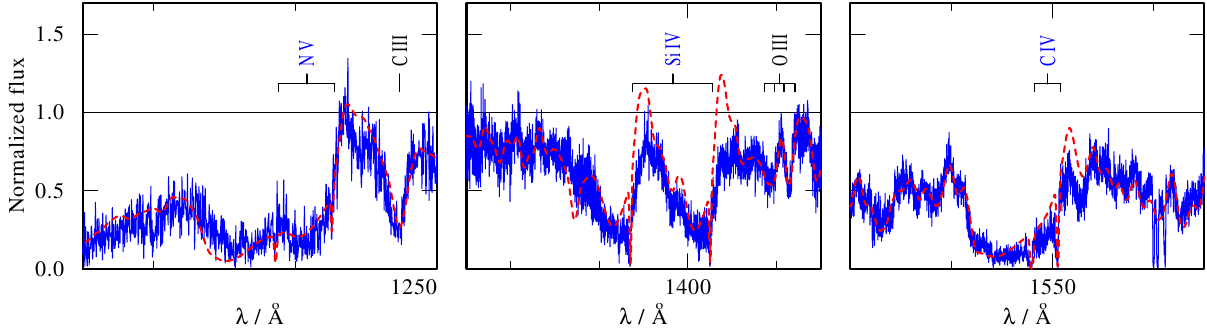}
    \includegraphics[width=1\linewidth]{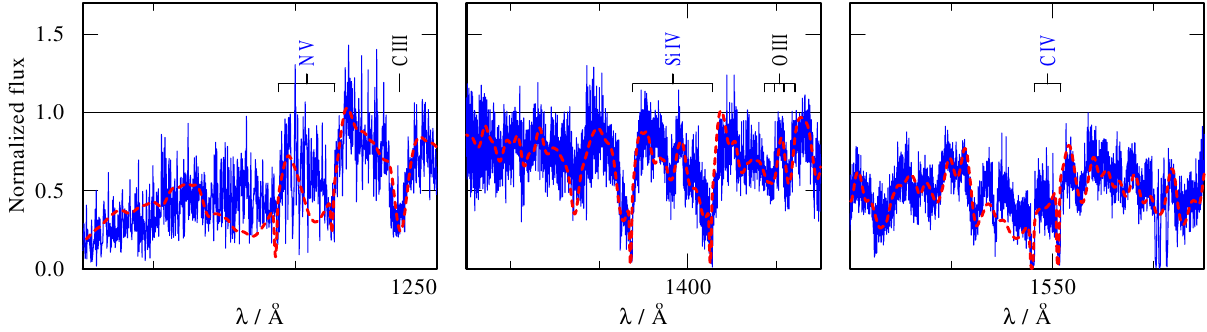}
    \caption{Comparison of UV P-Cygni profiles observed with HST (blue) at orbital phase $\phi \sim 0 $ (top panel) and $\phi \sim 0.5 $ (bottom panel) during high-soft state to the best-fit models (red). The models assume the same input parameters except for the incorporated amount of X-rays.
    The wind parameters used in the best-fit models are $\log \dot{M} = -6.5\,M_\odot \mathrm{yr}^{-1}$   and $\varv_{\infty}$ =1200\,km\,s$^{-1}$.
    For superior conjunction $\phi \sim 0 $ we adopted $L_X \sim 7 \times 10^{36}$\,erg\,s$^{-1}$ in the model, and at inferior conjunction $\phi \sim 0.5 $ we used higher amount of X-rays  $L_X \sim 2 \times 10^{37}$\,erg\,s$^{-1}$. The variability in the wind line profiles shows the impact of X-ray ionization.}
    \label{fig:uvlines}
\end{figure*}

\begin{figure*}
    \centering
    \includegraphics[width=1\linewidth]{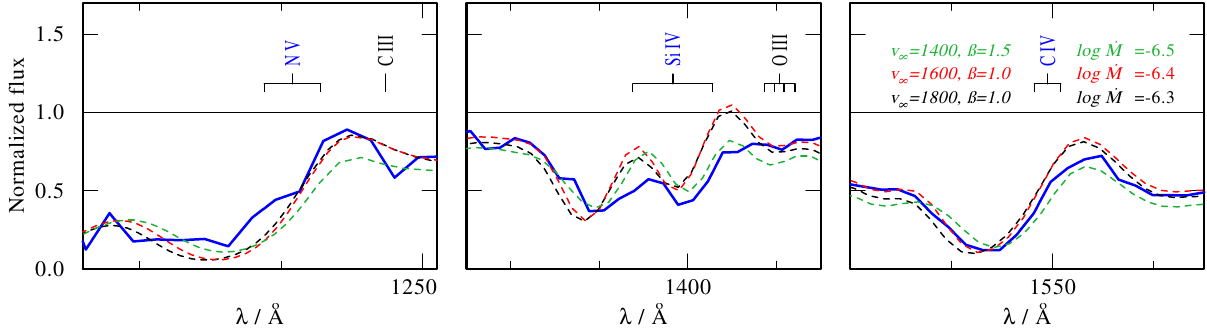} 
    \caption{Same as Fig.\,\ref{fig:uvlines} but IUE observations at orbital phase $\phi \sim 0 $ during low-hard state are shown.
    We adopted a lower X-ray luminosity $L_X \sim 3-5 \times 10^{36}$\,erg\,s$^{-1}$ in the model. Adopted wind parameters in models are shown in legend.}
    \label{fig:uvlines_iue}
\end{figure*}

\begin{table}[!hbt]
	\caption{Wind parameters derived for Cyg\,X-1 in this work at different X-ray states. }
	\label{table:windparameters}
	\centering
	\renewcommand{\arraystretch}{1.4}
	\begin{tabular}{lcc}
		\hline 
		\hline
		\vspace{0.1cm}
		Parameter                                         &          high-soft & low-hard       \\
		\hline
		\vspace{0.1cm}
        $\log \dot{M}$ ($M_\odot \mathrm{yr}^{-1}$)    & $-6.5^{+0.2}_{-0.2}$    & $-6.5$ to $-6.3$\\
		$\varv_{\infty}$ (km\,s$^{-1}$)     & $1200^{+100}_{-100}$  & 1400 to 1800  \\
            $\beta$                                        & $1.7^{+0.2}_{-0.2}$   &  1 to 1.5  \\  
		$D$                                            & $30^{+10}_{-10}$    &  $<30$   \\
		$\varv_\mathrm{wind}$ at BH (km\,s$^{-1}$)      & $341^{+50}_{-50}$ & 460 to 860\\
		\hline
	\end{tabular}
 
\end{table}

\subsection{Wind parameters}
Winds of massive stars are parameterized mainly by their mass-loss rate $\dot{M}$, terminal wind velocity $\varv_\infty$, velocity law exponent $\beta$, and clumping factor $D$. To determine these parameters using spectral analysis, in this work, we used UV spectra taken at $\phi\sim 0$ to yield parameters of the O star wind, while the spectra taken at $\phi\sim 0.5$  are employed to study the wind quenching by X-rays. The derived wind parameters at the high-soft  and low-hard states are listed in Table\,\ref{table:windparameters}.

The main diagnostics used to determine the terminal velocity is  the blue edge of prominent P\,Cygni lines such as  \ion{N} {v\,$\lambda\lambda$1238--1242},  \ion{Si}{iv} $\lambda\lambda$1393-1403 and \ion{C}{iv} $\lambda\lambda$1548-1551. The maximum blue edge velocity from the unsaturated P-Cygni lines is $\varv_\mathrm{edge}\approx 1300$\,km\,s$^{-1}$. A terminal velocity of $\varv_\infty \approx1200$\,km\,s$^{-1}$ shows better agreement with the observations in the high-soft state as shown in Fig.\,\ref{fig:uvlines}. This is in agreement with \cite{Gies2008}, where they used the SEI method to derive the wind velocity. On the other hand, the IUE spectra indicate higher wind velocities at low-hard state (Fig.\,\ref{fig:uvlines_iue}). While the low resolution of the IUE data precludes a precise measurement, the modeling of the \ion{C}{iv} profile suggests $\varv_\infty \approx 1600$\,km\,s$^{-1}$, and the \ion{N}{v} profile indicates an even higher value of $\varv_\infty \approx 1800$\,km\,s$^{-1}$. 
Even the hard-state values differ significantly from the commonly used value of $\approx2100$\,km\,s$^{-1}$, which was originally adopted in \citet{Herrero1992} based solely on optical spectra. \citet{Davis1983} also reported a value of 2300\,km\,s$^{-1}$, but this was based on examining IUE line widths alone, without any modeling.
With our obtained terminal velocity, the ratio of $\varv_\infty$ to the escape velocity of the donor at high-soft state is found to be $\varv_\infty/\varv_{\mathrm{esc}}\approx1.7$,  lower than the average value for typical O stars
\citep{Lamers1995,Kudritzki2000}. In contrast, the low-hard state exhibits a ratio of  $\lesssim 2.6 $, consistent with typical O star values.

For our spectral analysis, we computed models with $\log \dot{M}$ in the range of -6.0 to -7.0. To model the supersonic part of the wind, we tested $\beta$ values in the range 0.8 to 3. Since there are no simultaneous UV and X-ray observations of the system,  we varied $L_X \sim 4-9 \times 10^{36}$\,erg\,s$^{-1}$ in the model to match UV spectra taken at $\phi \sim 0 $ in the high-soft state. The P\,Cygni line profile shapes are best reproduced with $\beta=1.7$. Our best-fit model for the high-soft state incorporates X-rays with $L_X \sim 7 \times 10^{36}$\,erg\,s$^{-1}$ (see top panel in Fig.\,\ref{fig:uvlines}) and yields $\log \dot{M} =-6.5$ or $\dot{M} \approx 3\times 10^{-7}\,M_\odot \mathrm{yr}^{-1} $.
$\log \dot{M}$ between $-6.6$ and $-6.4$ reasonably reproduce the observed UV P\,Cygni profiles when varying $L_X  \sim 4 -9 \times 10^{36}$\,erg\,s$^{-1}$.
In contrast, to reproduce the IUE spectra at $\phi \sim 0 $ and in the low-hard state, models with slightly higher mass-loss rates ($\log \dot{M}$  between $-6.5$ and $-6.3$) and lower beta ($\beta<1.5$) are required (see Fig.\,\ref{fig:uvlines_iue}). To model the low-hard state, we use lower $L_X$ values in the range of $\sim 5-9 \times 10^{36}$\,erg\,s$^{-1}$.

One noticeable feature in the UV spectra are the weak emission components in  \ion{Si}{iv} $\lambda\lambda$1393-1403 and \ion{C}{iv} $\lambda\lambda$1548-1551, whereas OB supergiants with similar spectral types typically show \ion{Si}{iv} and \ion{C}{iv} resonance lines with strong emission in the red part. This is likely due to the asymmetries in the spatial distribution of corresponding ions destroyed by photoionization in the hemisphere towards the X-ray source. At phase 0 (BH behind the donor), the less-disturbed wind along our line of sight contributes to strong blue-shifted absorption in the P Cygni profiles, while X-ray photoionization on the opposite side weakens the emission component (as seen in top panel in Fig.\,\ref{fig:uvlines}).  If a focused wind component were present toward the BH, we would expect enhanced emission, particularly at phase 0.  This enhancement is not observed; the emission is notably weak. A similar absence of a stronger emission component is also observed in other BH-HMXBs such as M33 X-7 \citep{ramachandran2022}, while systems with neutron stars such as Vela X-1 or 4U 1700-37 show this emission component \citep[e.g.,][]{Sander2018,Hainich2020}  possible because of the lower X-ray luminosities of these systems.

The Hatchett-McCray effect is clearly observed in Cyg X-1 donor's UV wind lines.  X-ray photoionization strongly affects the wind on the X-ray facing hemisphere (phase 0.5), leading to higher ionization stages and decreased UV line intensity (Figure \ref{fig:uvlines}, bottom panel), compared to the less ionized wind observed at phase 0. The dramatic decrease in line strength at phase 0.5, especially at high velocities (Figure \ref{fig:uvlines}, upper and lower panels), indicates significant disruption of the donor wind by the accreting black hole and suggests that X-ray ionization extends to the photosphere. This decrease contradicts the expectation of increased absorption from a focused wind when the BH is in the line of sight \citep[e.g.,][]{Gies_Bolton1986,Sowers1998,Tarasov2003,Miller2005}, as discussed by \citet{Vrtilek2008}. Furthermore, our final model, using the same wind and stellar parameters \citep[except for an increased $L_X \sim 2 \times 10^{37}$\,erg\,s$^{-1}$  consistent with observations, e.g.,][]{Wilms2006,Sugimoto2017}  successfully reproduces the UV spectra at phase  $\phi \sim 0.5 $ (see bottom panel of Fig.\,\ref{fig:uvlines}). Therefore, UV observations show no evidence of increased mass flow focused towards the BH, further strengthening the argument against a focused wind.

A microturbulent velocity of 22\,km\,s$^{-1}$ in the photosphere is derived based on the optical lines, and is also in agreement with the UV spectra. Our adopted micro-turbulent velocity $\xi$ in the formal integral grows proportional to the wind velocity. 
For typical OB star analyses, it is sufficient to put the scale factor such that the maximum is $\xi(R_\mathrm{max})=0.1\varv_\infty$. However, for \mbox{Cyg\,X-1} we measure a higher turbulence velocity,
with a ratio $\xi(R_\mathrm{max})/\varv_\infty \approx$ 0.3, in agreement with the reported values for M33\,X-7 \citep{ramachandran2022}. The wind micro-turbulent velocities derived for these HMXBs are systematically larger compared those found in previous studies of OB stars in the Galaxy (e.g., 0.1 $\varv_\infty$ by \citealt{Kudritzki2000}, 0.14 $\varv_\infty$ by \citealt{Herrero2001}).  

The adopted microclumping is depth-dependent. Assuming a (maximum) clumping factor of $D=30$ and $R_{\mathrm{D}}=2R_\ast$ best reproduce the emission in \ion{He}{ii} at 4686\AA. Our analysis suggested that the clumping factor reaches this maximum value close to the location of the BH (2.1\,$R_\ast$).

In this work, we derive donor mass-loss rates of $\dot{M} \approx 3\times 10^{-7}\,M_\odot \mathrm{yr}^{-1} $ in the high-soft state while in the low-hard state we measured $\dot{M} \lesssim 5\times 10^{-7}\,M_\odot \mathrm{yr}^{-1} $. These significantly lower estimates represent a more detailed and accurate determination of the mass-loss rate compared to previous works.
For instance, \citet{Gies2003wind} derived $\dot{M} \approx 2-2.57\times 10^{-6}\,M_\odot\,\mathrm{yr}^{-1}$ across different X-ray states. However, their analysis was solely based on the H$\alpha$ equivalent width, neglected the effects of clumping and X-ray ionization, and relied on stellar parameters from \citet{Herrero1995}. Using the UV spectra taken during the high-soft state, \citet{Vrtilek2008} found a much higher mass-loss rate, $\dot{M} \approx 4.8\times 10^{-6}\,M_\odot\,\mathrm{yr}^{-1}$, using the SEI method.

Considering theoretical predictions, 
our derived wind mass-loss rate is a factor of $1.7-2.7$ smaller than predicted by the \citet{Vink2000} recipe for the same mass, luminosity, and temperature of the star at solar metallicity. When take into account the derived high iron abundances, the emprically derived  $\dot{M}$ is a factor of 2.2--3.5 lower than the ``Vink mass-loss rate'' at $Z=1.33Z_\odot$. This discrepancy can be attributed to the effects of wind clumping. Inadequate consideration of wind clumping can result in substantial overestimations of mass-loss rates, with discrepancies ranging from factors of 2 to 10 \citep[see, e.g.,][]{Puls2008,Fullerton2006}. This aligns with a recent analysis of early-type O supergiants in the Galaxy by \citet{hawcroft2021MW}, who reported a mass-loss rate reduction by a factor of $3.6$ compared to \citet{Vink2000} predictions, along with high clumping factors.

\subsection{Wind accretion versus\ Roche-lobe overflow}\label{sec:windvsroche}

In wind-fed supergiant-HMXBs, X-rays are created by the accretion of wind material from the donor star onto the compact companion. The mass accretion rate via the Bondi-Hoyle formula is 
\begin{equation}
  \label{eq:Mdotacc}
\dot{M}_\mathrm{acc} =\pi	R_\mathrm{acc}^{2} \rho \varv_{\mathrm{rel}},
\end{equation} where $\rho$ is the wind density near the BH and $R_\mathrm{acc}$ is the accretion radius. 
In \mbox{Cyg\,X-1}, the wind velocity near the BH (at 2.1$R_\ast$) reaches around $\approx340$\,km\,s$^{-1}$ at high-soft state and up to $\lesssim 860$\,km\,s$^{-1}$ at low-hard state. 
The value of the velocity exponent $\beta$ derived in this work ranges from 1 to 1.7, defining the wind structure and its influence on the wind velocity determined for the position of the BH. This higher $\beta$ value indicates a more gradual acceleration of the wind, resulting in lower wind velocities near the BH orbit compared to the typically assumed value of $\beta=0.8$, which would imply a steeper acceleration and higher wind velocities. 
The mass accretion rate onto the compact object is a sensitive function of the wind velocity law, following the relationship $\dot{M}_\mathrm{acc} \propto \varv^{-4}$. This means that even small changes in the wind velocity can lead to significant variations in the mass accretion rate \citep[e.g.,][]{Oskinova2012}.

The orbital velocity ($2\pi a/P$) near the BH is $\approx270$\,km\,s$^{-1}$. The relative velocity between the accreting BH and the stellar wind can be estimated using  $\varv_{\mathrm{rel}}^{2} = \varv_{\mathrm{wind}}^{2} + \varv_{\mathrm{orbit}}^{2}    \approx 435$\,km\,s$^{-1}$ at high-soft state. This derived $\varv_{\mathrm{rel}}$ is only $0.36\,\varv_\infty $. In the low-hard state, $\varv_\mathrm{rel} \lesssim 901 \, \mathrm{km/s}$  or $\sim 0.5\,\varv_\infty $.
The density of the wind at the distance of the BH can be approximated to 
\begin{equation}
	\rho = \dfrac{\dot{M}_\mathrm{donor}}{4\pi d_\mathrm{BH}^{2}  \varv_{\mathrm{wind}} }.
\end{equation}
By inserting the derived wind parameters, we get a density of $\rho \approx 4\times 10^{-15} \mathrm{g\,cm^{-3}}$ in the high-soft state and $\gtrsim 1.5 \times 10^{-15} \, \mathrm{g/cm^3}$ in the low-hard state.
The accretion radius, $R_\mathrm{acc}= 2GM_\mathrm{BH}/\varv_\mathrm{rel}^{2}$, is $2.4\times 10^{7}$\,km  in the high-soft state, yielding a mass accretion rate of $\dot{M}_\mathrm{acc}=3.3\times 10^{18}\,\mathrm{g\,s^{-1}}$.  In the low-hard state, the smaller accretion radius, $R_\mathrm{acc} \approx 6\times 10^{6}$\,km, results in a lower mass accretion rate, $\dot{M}_\mathrm{acc} \approx 4.4 \times 10^{17}\,\mathrm{g\,s^{-1}}$.  

The accretion disk luminosity in BHs arises from material spiraling inward, releasing gravitational energy as it moves from the outermost radius to the innermost stable circular orbit $R_\mathrm{BH}$. The accretion luminosity can be written as
\begin{equation}\label{eq:Lacc}
L_{\mathrm{acc}} = \dfrac{GM_\mathrm{BH}\dot{M}_\mathrm{acc}}{R_\mathrm{BH}} = \eta \dot{M}_\mathrm{acc} \text{.}
\end{equation}

Rapidly spinning black holes, like the one in Cyg X-1, have accretion disks with inner edges much closer to the event horizon. This significantly increases the accretion efficiency $\eta$ and hence the $L_{\mathrm{acc}}$. For Cyg X-1, estimates of the spin parameter range from $a_\ast\sim 0.9$ \citep{zdziarski2024spin} to nearly 1 \citep{zhao2021spin}. Our lower BH mass estimate suggests a potentially slightly lower spin ($a_\ast\sim 0.7-0.8 $), but high efficiency remains likely.  For instance, with $\eta=0.2$ the resulting accretion luminosity would be \(L_{\mathrm{acc}} \sim 6 \times 10^{38} \, \mathrm{erg\,s^{-1}}\) in the high-soft state and \(L_{\mathrm{acc}} \sim 4.5 \times 10^{37} \, \mathrm{erg\,s^{-1}}\) in the low-hard state.

Although the calculated accretion luminosity is below the Eddington limit \(L_{\mathrm{Edd}} \sim 2 \times 10^{39} \, \mathrm{erg\,s^{-1}}\), it significantly exceeds the observed X-ray luminosity -- by an order of magnitude in the high-soft state and a factor of 2 in the low-hard state.
Such large differences between Bondi-Hoyle accretion luminosity and observed X-ray luminosity have been reported for other BH HMXB systems such as IC\,10\,X-1 and NGC\,300\,X-1 \citep{Tutukov2016ARep}. This discrepancy highlights the limitations of the Bondi-Hoyle accretion model, which simplifies the complex physics of accretion.  Our analysis also uses a typical $\beta$-law assumption for the wind velocity field, potentially underestimating the wind velocity near the BH. Moreover, in the above calculations we used the BH mass close to its maximum value of $17.5\,M_\odot$. Variations in accretion efficiency, radiative processes, absorption, scattering, and geometric effects can further contribute to discrepancies between theoretical and observed luminosities.

Our calculations indicate that the observed X-ray luminosity in \mbox{Cyg\,X-1} could be fully explained by the accretion of the donor star's stellar wind. If there would be additional mass flow in the direction of the BH due to the Roche lobe overflow of the donor, this should cause a much higher X-ray luminosity than currently observed. The latter phenomenon is observed in systems like M33 X-7, where the stellar wind of the donor is insufficient to account for the high X-ray luminosity, indicating the need for invoking additional mass flow towards the BH by Roche-lobe overflow or wind-RLOF   \citep{ramachandran2022}. 

As we discuss and illustrate in Sect.\,\ref{sec:mass}, in \mbox{Cyg\,X-1} the donor radius matches or even exceeds the equivalent radius of the Roche lobe. However, the conventional Roche lobe definition might not be applicable in a straightforward way to an O-type donor star with radiatively driven stellar wind.  
While we do not solve the wind hydrodynamics in this work and thus cannot examine the acceleration in the launching regime in more detail, the presence of a radiation-driven wind implies $\Gamma_\text{rad} > 1$, meaning that a conventional Roche lobe actually does not exist for the donor \citep[see, e.g.,][]{Dermine+2009}.   
Even in the case of strong X-ray illumination, our models show $\Gamma_\text{rad} \approx 1$ at the radius corresponding to the orbit of the BH. The acceleration in the outer wind is much higher than this. In addition, we see evidence of wind signatures in the UV spectrum during $\phi \sim 0.5 $ such as \ion{C}{iv}  and \ion{N}{v} resonance lines (see Fig.\,\ref{fig:uvlines}). We thus conclude from our modeling that there is no conventional Roche lobe overflow present in \mbox{Cyg\,X-1} which is in line with our finding that the stellar radius does not yet reach the radius of the inner Lagrangian point (cf.\ Fig.\,\ref{fig:roche-geo}).

Although the precise drivers of \mbox{Cyg\,X-1}'s X-ray state transitions remain unclear, our wind analysis points to a possible connection with wind-accretion feedback. In the high-soft state, the lower wind velocity and mass-loss rate result in a higher mass accretion rate and, consequently, a higher X-ray luminosity. Conversely, in the low-hard state, the significantly higher wind velocity and moderately higher mass-loss rate result in a lower mass accretion rate and, as a result, lower X-ray luminosity. This suggests a complex interplay where changes in the wind affect the accretion flow, which in turn influences the observed state transitions.  Further investigation is needed to disentangle this complex relationship.

\section{Evolution of the system}
 
We explore the system's future evolution channel towards binary BHs and investigate its possibility of becoming a gravitational wave source. We use the stellar evolution code MESA (``Modules for Experiments in Stellar Astrophysics'', Version No.\ 10398) with a physics implementation as described in \citet{Paxton2011,Paxton2013,Paxton2015,Paxton2018} to compute the tracks. 
Given the substantial uncertainties in the system's prior evolution (e.g., stable mass transfer, common envelope evolution, supernova kick uncertainties, potential dynamical capture), modeling it is beyond the scope of this work. Our focus is to assess the system's current evolutionary state—specifically, whether it is wind-fed or currently undergoing Roche lobe overflow. We model the binary evolution, starting from a configuration with a ZAMS O-star donor (initial mass in the range of 29--35~$M_\odot$) and a BH companion (17.4~$M_\odot$) and a 5.5~day initial period. These initial parameters are chosen so that they are close to the estimated masses and period of the system from observations, and the main objective is to match the current state of the system. Evolution is then computed until either core carbon is depleted or until a merger between the
donor star and the BH occurred. After core carbon depletion, a star is assumed to directly collapse into a BH without losing mass. Subsequently, we assign the final core mass to that of the compact companion.

Our models are computed with a solar metallicity of $Z = 0.0142$ \citep{Asplund2009}, where we take the relative metal mass fractions from \citet{Grevesse1998}. In our calculation, the binary system is assumed to be synchronized with the orbital period at the ZAMS. The evolution of orbital angular momentum takes into account spin-orbit coupling, gravitational wave radiation, and mass loss, as explained by \citet{Paxton2015}. 

Convection was modeled using the standard mixing-length theory \citet{BohmVitense1958} with a mixing-length parameter $\alpha = 1.5$, adopting the
Ledoux criterion. Convective overshooting is implemented with a step-overshooting scheme, extending the convective core by $\alpha_{\text{ov}} = 0.345$ pressure scale heights, thereby following \citet{Brott2011}. The efficiency parameter $\alpha_\mathrm{SC} = 1.0$ is used to represent semiconvection, following the work of \citet{Langer1983}.
The effect of the centrifugal force was implemented as in \citet{Heger2000}. Rotation-induced composition-and angular-momentum transfer encompasses the GSF instability, secular and dynamical shear instabilities, and the impacts of Eddington-Sweet circulation.
 
We use MESA's ``Dutch'' scheme for wind predictions, with cool star winds following \citet{Nieuwenhuijzen1990}, hot-star winds applying \citet{Vink2000,Vink2001}, and Wolf-Rayet stars following \citet{Nugis2000}. The mass-loss rates in these prescriptions are most likely an overestimation, for example due to their assumptions of homogeneous winds. Our analysis of the donor wind suggests a high clumping factor and lower mass-loss rate than predicted by \citet{Vink2001} for the same stellar parameters. We thus computed models with scaling factors of $1$, $0.5$, and $0.2$ for the wind mass loss in order to approximately counteract a possible overestimation of mass-loss rates.
 
\begin{figure}
    \centering
    \includegraphics[width=1\linewidth]{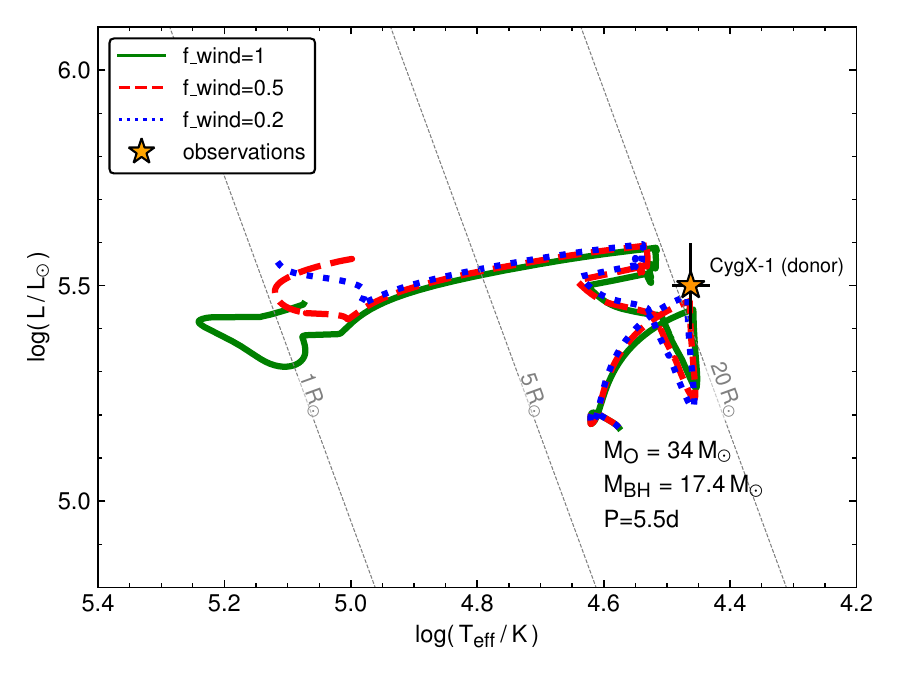}
    \caption{Potential future evolution of the \mbox{Cyg\,X-1} donor star. Here, the evolution starts when the primary is already a BH (point mass) of 17.4\,$M_\odot$,  O star donor of initial mass 34\,$M_\odot$, and an initial period of 5.5 days. Three models are displayed here with mass-loss scaling factors of 1, 0.5, and 0.2 to demonstrate the effect of wind mass loss on evolution.  }
    \label{fig:HRD-BH}
\end{figure}

To find the best matching models for the observed system parameters, we compare the stellar parameters of the O star donor derived from spectral analysis -- such as effective temperature, luminosity, and surface gravity -- along with the current orbital period of the system, to the evolutionary model tracks. Models starting with an O star donor of 34\,$M_\odot$ and a BH companion of 17.4\,$M_\odot$ with a $5.5$-day period successfully reproduce the observed properties of the \mbox{Cyg\,X-1} system within uncertainties for all three wind mass-loss scaling factors. The corresponding tracks are shown in the HRD in Fig.\,\ref{fig:HRD-BH}, and the best-matching parameters from these models are compared in Table\,\ref{table:best-match}. The current wind mass-loss rate of the O star donor is best reproduced by the evolutionary model that assumes a mass loss scaling factor of 0.5 (model ``M2'' in Table \ref{table:best-match}).

\begin{table}[h!]
\centering
\caption{Parameters of MESA Models with different mass-loss scaling factors compared to the observed values of the donor in \mbox{Cyg\,X-1}. }
\label{table:best-match}
\renewcommand{\arraystretch}{1.4} 
\begin{tabular}{lllll}
\hline
\hline
Parameter & M1 & M2 & M3 & Observation \\
 $f_{\mathrm{wind}}$&1&$0.5$&$0.2$&\\
\hline
$T_{\rm{eff}}$ (kK) &    28.6  & 29.1 & 29.3 & $28.5^{+1}_{-1}$\\
$\log L$ ($L_\odot$) &    5.41   & 5.46 & 5.47 & $5.5^{+0.1}_{-0.1}$\\
$\log g$ &    3.29   & 3.29 & 3.30& $3.2^{+0.1}_{-0.1}$  \\
$R$ ($R_\odot$)  &      20.58   & 21.21 & 21.20&$22.3^{+1.5}_{-2.5}$ \\
$M_\mathrm{donor}$ ($M_\odot$)  & 30.4   & 32.6 & 33.4 & $29^{+6}_{-3}$\\
$M_\mathrm{BH}$ ($M_\odot$) &    17.58  & 17.49 & 17.43 & $17.5^{+2}_{-1}$\\
$P_{\rm{orb}}$ (day) &       5.589  & 5.590 & 5.501 &5.599\\
$f_\mathrm{RL}$ &0.99 &0.99 &0.99 & $1.1^{+0.15}_{-0.2}$\\
$\log \dot{M}$ ($M_\odot\,\mathrm{yr}^{-1}$) & -6.34  & -6.49 & -6.85 & $-6.5^{+0.2}_{-0.2}$\\
\hline
\end{tabular}
\end{table}

\begin{figure*}
    \centering
    \includegraphics[width=0.5\linewidth]{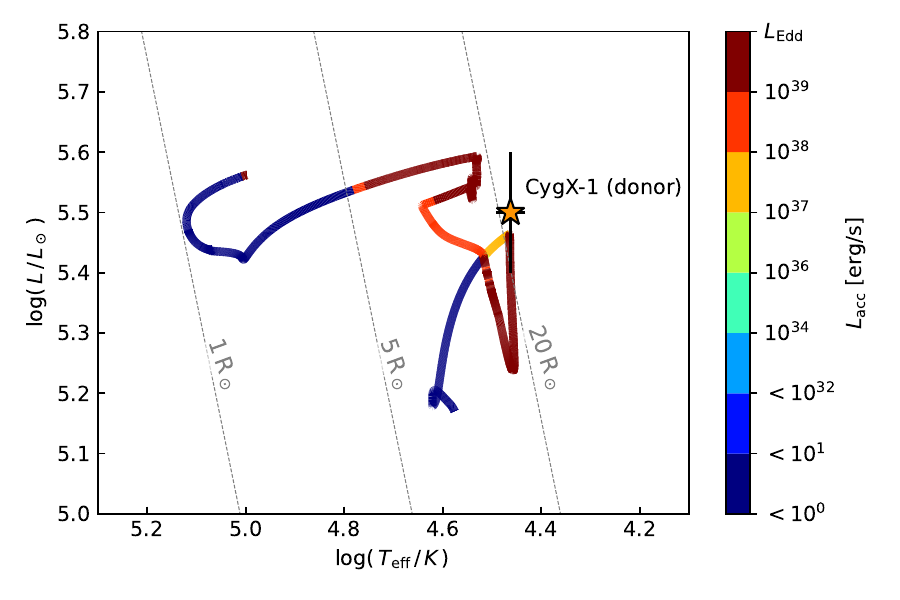}
    \includegraphics[width=0.4\linewidth]{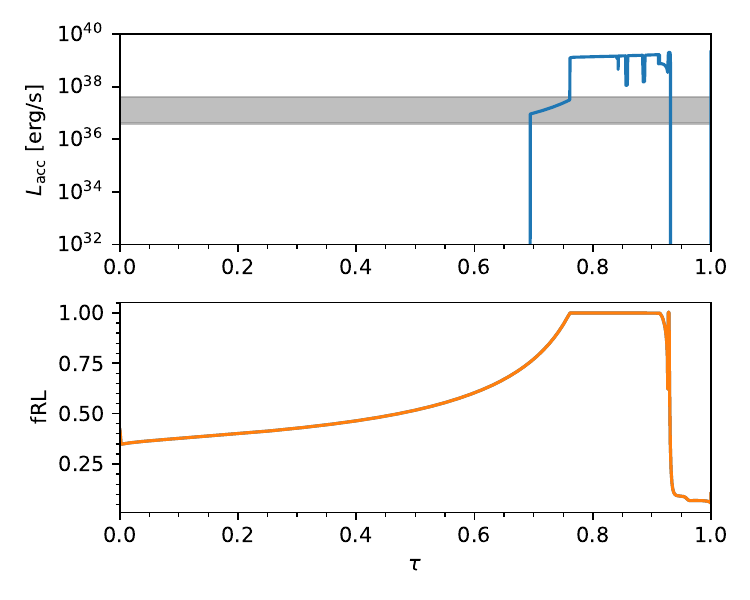}
    \caption{ Left panel: Same as Fig.\,\ref{fig:HRD-BH} but here the tracks correspond to a wind mass-loss scaling factor of 0.5, and the color scaling shows the accretion luminosity. Right panel: Accretion luminosity (top) and Roche-lobe filling factor (bottom) as a function of normalized donor lifetime $\tau$. The gray bar shows the range of observed X-ray luminosity of the system. }
    \label{fig:Lacc}
\end{figure*}

\begin{figure}
    \centering
    \includegraphics[width=0.9\linewidth]{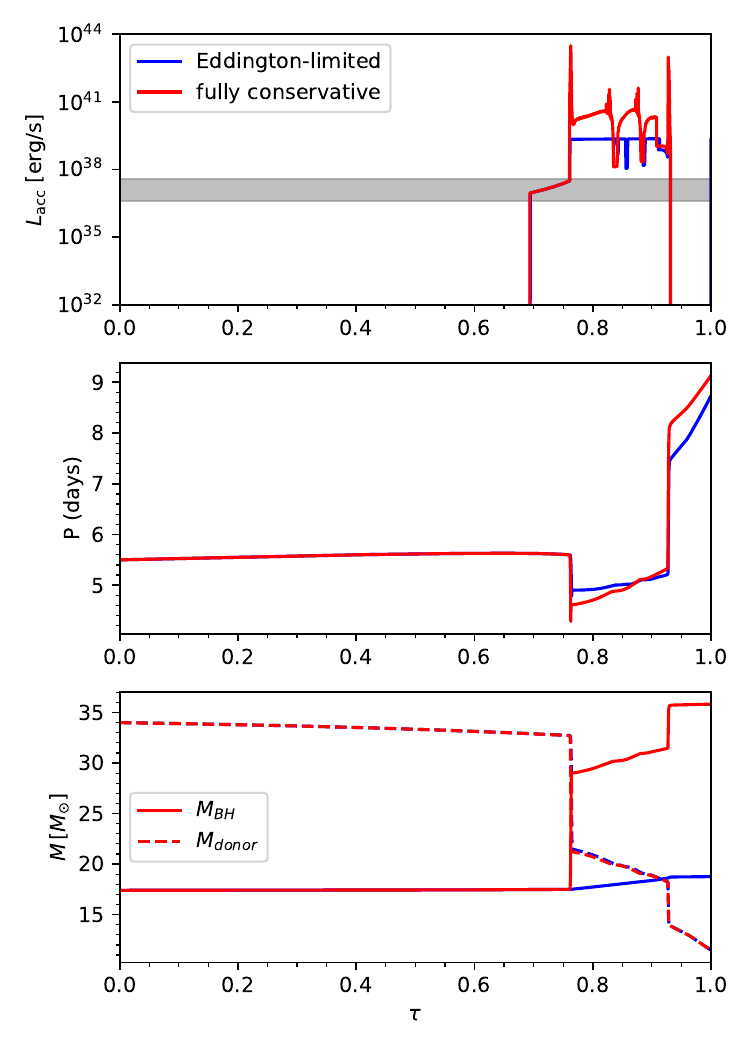}
    \caption{Impact of mass accretion by the BH on the evolution of Cyg X-1. The panels show the evolution of accretion luminosity, orbital period, and component masses as a function of normalized time $\tau$ for MESA models with Eddington-limited (blue) and fully conservative (red) mass accretion.}
    \label{fig:BHaccretion}
\end{figure}

The current properties of the system match with a phase just before the first mass transfer from the donor to the BH via RLOF.
At this stage, the donor star is still burning hydrogen in its core, indicating it is on the main sequence or slightly evolved from it.  
After the mass transfer, the donor's mass is reduced to approximately 
$21\,M_\odot$ and continues hydrogen burning for another $\sim1$ million year. Despite this reduction in mass and luminosity, the donor retains the appearance of an O supergiant, along with surface He and N enrichment.
When hydrogen is depleted in the core and helium burning begins, the next mass transfer phase is initiated. During this phase, the donor's mass decreases further, eventually falling below  $14\,M_\odot$. The donor star rapidly contracts after that, becoming a Wolf-Rayet (WR) star.

The stellar wind during the WR phase is significantly influenced by the different mass-loss scaling factors adopted in our evolutionary models. For instance, a mass-loss scaling factor (\( f_{\text{wind}} \)) of $1$ and $0.5$ leads to more extensive stripping of the stellar layers, driving the star to a more evolved WC phase. In contrast, a model with a lower mass-loss rate \( f_{\text{wind}} = 0.2 \) remains in the WN phase. Additionally, the orbit of the binary system widens due to the strong wind mass loss during the WR phase. The final core masses of the donor range from 9.5 to $13.3\,M_{\odot}$, and the final orbital period varies from 10.9 to 7.5 days for \( f_{\text{wind}} = 1 \) to \( f_{\text{wind}} = 0.2 \).

We calculated the accretion luminosity throughout the system's evolution using Eq.\,\eqref{eq:Lacc}, accounting for both Roche lobe overflow and wind mass transfer (Fig.\,\ref{fig:Lacc}). For $\varv_{\mathrm{wind}}$, we assumed a $\beta$ law with a standard $\beta$-value of 1.  The terminal velocity $\varv_\infty$ can be estimated from the effective escape velocity
\begin{equation}
  \varv_{\mathrm{esc},\Gamma} = \sqrt{\frac{2 G M}{R} \left(1 - \Gamma_\text{e}\right)}\text{.}
\end{equation}
Empirical studies of OB stars in the Milky Way suggest $\varv_\infty$/$\varv_{\mathrm{esc},\Gamma}\simeq$\,2.6 for stars with $T_\ast\geq$\,21\,kK \citep{Lamers1995,Kudritzki2000}. A similar ratio is also found for Galactic WN stars by \citet{Niedzielski2002}. The accretion luminosity is calculated only when an accretion disk formed around the BH, requiring the accreted material to have sufficient angular momentum to exceed that of the innermost stable circular orbit \citep[see equation 10 in][]{Sen2021}
Mass transfer is treated using the 'roche lobe' scheme. In this scheme, the mass transfer rate is dynamically adjusted to ensure the donor star remains within its Roche lobe, with the Roche-lobe radius computed as in \citet{Eggleton1983}. The observed X-ray luminosity matches the accretion luminosity of the evolutionary model just before the mass transfer phase (yellow region in Fig. \ref{fig:Lacc}). Soon the system will undergo RLOF and the luminosity will be increased $\gtrsim10^{39}$ erg\,s$^{-1}$. The donor star continues to fill its Roche lobe until it finishes the second mass transfer phase and contracts towards the WR stage. During this time, the system will appear as an ultra-luminous X-ray source (ULX).

Initially, we assumed a highly non-conservative mass accretion scenario where the BH accretes at the Eddington limit, with the remaining material ejected from the system carrying the specific angular momentum of the accretor. This resulted in limited BH growth and consequently, wide post-mass transfer orbits. As a result, the predicted merger timescale is significantly longer, estimated to be approximately 51 Gyr. Three BH accretion scenarios -- Eddington-limited, moderate super-Eddington, and totally conservative accretion -- have been compared in a recent work by \citet{Xing2025}. Considering their findings, the only scenario that could result in a potential BH merger for \mbox{Cyg\,X-1} would be a highly conservative accretion scenario.

Motivated by this, we re-examined our best-fit model (M2, with \( f_{\text{wind}} = 0.5 \)) using a fully conservative accretion scenario. While this change does not significantly affect the currently observed parameters of \mbox{Cyg\,X-1} (which match a pre-Roche-lobe overflow state, see Table \ref{table:best-match}), it drastically alters the future evolutionary outcome. As shown in Fig. \ref{fig:BHaccretion}, the final BH mass increases substantially (from 18.7 to $35.8\,M_\odot$) due to the enhanced accretion. Enhanced accretion would also result in much higher X-ray luminosities on the order of $\gtrsim10^{40-43}$ erg\,s$^{-1}$ during or after the mass transfer phase.  The period of the system only slightly increases, and the mass of the donor remains the same. Moreover, \citet{Xing2025} suggest that only a highly conservative accretion onto BHs lead to extreme BH spins \(a_\ast \sim 1\). Our fully conservative model suggests that the system would become binary BHs with final masses of  $\sim 35.8\,M_\odot$ and  $\sim 11.5\,M_\odot$ and a final period of $\sim$$9.13$ days, making it a strong gravitational wave progenitor that will merge in approximately 5\,Gyr. It should be noted that this merger timescale will increase if we consider lower values for the current BH mass in the model. Nonetheless, our findings indicate that mass accretion efficiency plays a crucial role in the evolution of binary systems toward becoming merging black holes and potential gravitational wave sources, with Cyg X-1 as a prime example.

\section{Summary}

We have presented a comprehensive analysis of \mbox{Cyg\,X-1},  utilizing archival high-resolution UV and optical spectra in conjunction with sophisticated atmospheric models. Notably, this is the first investigation to simultaneously analyze UV and optical spectra of \mbox{Cyg\,X-1} while also incorporating X-rays to constrain the stellar and wind parameters. Our key findings are as follows:

\begin{itemize}
    \item Our analysis yields notably lower masses for both the donor ($\approx 29\,M_\odot$) and the BH. Assuming the inclination from the literature, we obtain a BH mass of $\approx 17.5\,M_\odot$, which is most likely an upper limit. Estimating the inclination from tidal synchronization yields a BH mass of $\approx 13.8\,M_\odot$, which is also close to the peak mass  $\approx 14.2\,M_\odot$  predicted by our Monte Carlo simulation probing a range of inclinations. 
    
    \item We found evidence for N and He enrichment at the surface of the donor. This is likely a result of the combined effects of material transfer from the BH progenitor and subsequent efficient mixing driven by rapid rotation.

    \item The analysis reveals that \mbox{Cyg\,X-1} exhibits super-solar abundances of Fe, Si, and Mg, approximately 1.3--1.8 times higher than solar values. Notably, while there is evidence of CNO-processed material on the surface (nitrogen enrichment, oxygen depletion), the total CNO abundance remains consistent with solar values. This distinct abundance pattern sets \mbox{Cyg\,X-1} apart from stars in its surrounding Cyg OB3 association.

    \item While a supernova from the BH progenitor could potentially enrich the donor star, the observed abundance pattern (specifically the simultaneous enrichment of Fe, Si, and Mg without a corresponding increase in total CNO) is difficult to reconcile with current models of supernova yields and pre-supernova wind yields. This implies that the system had a high initial metallicity and likely formed from an enriched ISM.

   \item Our analysis yields a terminal wind velocity of $\approx 1200\,\mathrm{km\,s}^{-1}$ and a  mass-loss rate of $\dot{M} \approx 3\times 10^{-7}\,M_\odot\,\mathrm{yr}^{-1}$ in the high-soft state. We found evidence for faster and stronger wind in low-hard state ($\varv_\infty \lesssim 1800\,\mathrm{km\,s}^{-1}$ and $\dot{M} \lesssim 5\times 10^{-7}\,M_\odot\,\mathrm{yr}^{-1}$). These estimates are significantly lower compared to the commonly adopted values in the literature for this system. This reduction is primarily attributed to several key improvements in our modeling approach: (1) the inclusion of X-ray photoionization effects, (2) the use of advanced, non-LTE atmospheric models with stellar winds (3) the simultaneous analysis of UV and optical wind features, (4)  consideration of wind clumping, and (5) a detailed exploration of the wind velocity exponent. This derived mass-loss rate is lower than \citet{Vink2000} predictions by a factor of $\approx2-3$, further emphasizing the impact of clumping.

    \item The  X-ray luminosity of \mbox{Cyg\,X-1} is consistent with being driven primarily by wind accretion. Our Bondi-Hoyle calculations predict a higher accretion luminosity than observed, suggesting that not all accreted material contributes to the X-ray emission, and the persistent presence of wind features in the UV spectra, even under strong X-ray ionization, confirms the dominance of wind-driven mass transfer over RLOF. Furthermore, the UV observations are inconsistent with the presence of a focused wind toward the BH.
    
    \item We find that the radius of the donor star is larger than the equivalent spherical radius of its Roche lobe, adopting a classical description. However, the strong radiation force of the O-star may significantly change its effective potential and thereby modify the typical Roche lobe geometry.
     The donor star radius is smaller than the radius of the inner Lagrangian point for all plausible values of inclination and the BH mass. We thus conclude that there is no evidence for a significant focused direct overflow towards the black hole in \mbox{Cyg\,X-1}. 

    \item The observed variations in wind velocity and mass-loss rate, coupled with corresponding changes in X-ray luminosity between high-soft and low-hard states, point to a complex interplay between the stellar wind, accretion, and X-ray states in Cyg X-1.

    \item MESA evolutionary models constrained by newly derived stellar parameters show that \mbox{Cyg\,X-1} is currently in a pre-RLOF phase, consistent with the observed X-ray luminosity. However, these models predict that the system will undergo RLOF in the near future, leading to a significant increase in X-ray luminosity, potentially becoming a ULX. The subsequent evolution depends strongly on the wind mass-loss rate of the donor and mass accretion efficiency onto the BH.
    
     \item We consider Cyg\,X-1 as a potential gravitational wave source progenitor.  While  Eddington-limited mass accretion scenarios lead to limited BH growth and wider orbits (merging time $>51$\,Gyr), the models assuming fully conservative mass accretion predict a dramatic increase in the BH mass, a modest increase in the orbital period, and a final binary BH system that will merge within approximately $\gtrsim5$\,Gyr, making \mbox{Cyg\,X-1} a potential gravitational wave source. This highlights the critical role of mass accretion efficiency in determining the system's potential as a gravitational wave source.

\end{itemize}
Taken together, these results paint a complex and interconnected picture of the Cyg X-1 system, highlighting the crucial role of stellar winds, accretion processes, and initial conditions in shaping its evolution and ultimate fate.

\begin{acknowledgements}

VR, AACS, and MBP are supported by the Deutsche Forschungsgemeinschaft (DFG - German Research Foundation) in the form of an Emmy Noether Research Group -- Project-ID 445674056 (SA4064/1-1, PI Sander).  
ECS acknowledges financial support by the Federal Ministry for Economic Affairs and Climate Action (BMWK) via the German Aerospace Center (Deutsches Zentrum f\"ur Luft- und Raumfahrt, DLR) grant 50 OR 2306 (PI: Ramachandran/Sander).
DP acknowledges financial support from the Research Foundation Flanders (FWO) under grant agreement No. 1256225N.   B.K. gratefully acknowledges support from the Grant Agency of the Czech
Republic (GA\v CR 25-15910S). The Astronomical Institute of the Czech
Academy of Sciences in Ond\v rejov is supported by the project
RVO:67985815.
\end{acknowledgements}
 
\bibliographystyle{aa} % style aa.bst
\bibliography{ref} % your references Yourfile.bib

\begin{appendix}
% \appendix
\section{Additional tables}

\begin{table}[htbp]
\centering
\caption{Atomic model used to construct the PoWR models in this work.}
\label{tab:modelatom}
\begin{tabular}{lll|lll}
\hline %-------------------------------------------------------------------------------
\hline
Ion& Number of levels& Number of lines & Ion& Number of levels& Number of lines \\
\hline
\ion{H}{i} & 22 & 231 & \ion{Mg}{iii} & 10 & 45 \\
\ion{H}{ii} & 1 & 0 & \ion{Mg}{iv} & 1 & 0 \\
\ion{He}{i} & 35 & 595 & \ion{Si}{ii} & 1 & 0 \\
\ion{He}{ii} & 26 & 325 & \ion{Si}{iii} & 24 & 276 \\
\ion{He}{iii} & 1 & 0 & \ion{Si}{iv} & 23 & 253 \\
\ion{N}{i} & 10 & 45 & \ion{Si}{v} & 1 & 0 \\
\ion{N}{ii} & 38 & 703 & \ion{P}{iv} & 12 & 66 \\
\ion{N}{iii} & 36 & 630 & \ion{P}{v} & 11 & 55 \\
\ion{N}{iv} & 38 & 703 & \ion{P}{vi} & 1 & 0 \\
\ion{N}{v} & 20 & 190 & \ion{Al}{ii} & 10 & 45 \\
\ion{N}{vi} & 14 & 91 & \ion{Al}{iii} & 10 & 45 \\
\ion{C}{i} & 15 & 105 & \ion{Al}{iv} & 10 & 45 \\
\ion{C}{ii} & 32 & 496 & \ion{Al}{v} & 10 & 45 \\
\ion{C}{iii} & 40 & 780 & \ion{Ne}{i} & 8 & 28 \\
\ion{C}{iv} & 25 & 300 & \ion{Ne}{ii} & 1 & 0 \\
\ion{C}{v} & 5 & 10 & \ion{Ne}{iii} & 18 & 153 \\
\ion{C}{vi} & 15 & 105 & \ion{Ne}{iv} & 35 & 595 \\
\ion{O}{i} & 13 & 78 & \ion{Ne}{v} & 54 & 1431 \\
\ion{O}{ii} & 37 & 666 & \ion{Ne}{vi} & 49 & 1176 \\
\ion{O}{iii} & 33 & 528 & \ion{Ne}{vii} & 1 & 0 \\
\ion{O}{iv} & 25 & 300 & \ion{G}{i}\tablefootmark{*} & 1 & 0 \\
\ion{O}{v} & 36 & 630 & \ion{G}{ii}\tablefootmark{*} & 3 & 2 \\
\ion{O}{vi} & 16 & 120 & \ion{G}{iii}\tablefootmark{*} & 13 &40  \\
\ion{O}{vii} & 15 & 105 & \ion{G}{iv}\tablefootmark{*} & 18 & 77 \\
\ion{S}{iii} & 23 & 253 & \ion{G}{v}\tablefootmark{*} & 22 & 107 \\
\ion{S}{iv} & 11 & 55 & \ion{G}{vi}\tablefootmark{*} & 29 & 194 \\
\ion{S}{v} & 10 & 45 & \ion{G}{vii}\tablefootmark{*} & 19 &  87\\
\ion{S}{vi} & 1 & 0 & \ion{G}{viii}\tablefootmark{*} & 14 & 49 \\
\ion{Mg}{i} & 1 & 0 & \ion{G}{ix}\tablefootmark{*} & 15 & 56 \\
\ion{Mg}{ii} & 12 & 66 & \ion{G}{x}\tablefootmark{*} & 1 & 0 \\
\noalign{\vspace{1mm}}
\hline
\end{tabular}
\tablefoot{
\tablefoottext{*}{G denotes a generic atom which incorporates the following
iron group elements: Sc, Ti, V, Cr, Mn, Fe, Co, and Ni. The corresponding ions are treated by means of a superlevel approach (for details, see
\citet{Graefener2002}.} 
}
\end{table}

\begin{table}[]
\centering
\caption{Number fractions and mass fractions of iron group elements relative to Fe used in PoWR model calculation.}
\label{tab:Fegroup}
\begin{tabular}{llll}
\hline %-------------------------------------------------------------------------------
\hline%-------------------------------------------------------------------------
\noalign{\vspace{1mm}}
Element  & $Z$  & $n_i/n_{\rm Fe}$   & $X_i/X_{\rm Fe}$   \\
\hline
\noalign{\vspace{1mm}}
Sc      & 21 & 3.09E-05 & 2.49E-05 \\
Ti      & 22 & 2.40E-03 & 2.06E-03 \\
V       & 23 & 2.95E-04 & 2.69E-04 \\
Cr      & 24 & 1.29E-02 & 1.20E-02 \\
Mn      & 25 & 9.33E-03 & 9.18E-03 \\
Fe      & 26 & 1.00        & 1.00 \\
Co      & 27 & 2.24E-03 & 2.36E-03 \\
Ni      & 28 & 4.79E-02 & 5.03E-02\\
\hline
\end{tabular}
\end{table}

\section{Additional plots}

\begin{figure*}
    \centering
    \includegraphics[width=1.\linewidth]{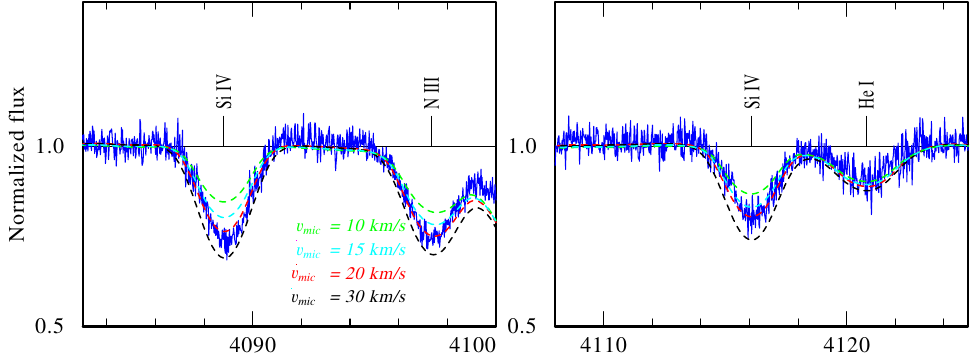}
    \caption{Impact of $\xi$ on metal line profiles.}
    \label{fig:vmic}    
\end{figure*}

\begin{figure*}
    \centering
    \includegraphics[width=1.\linewidth,angle=90]{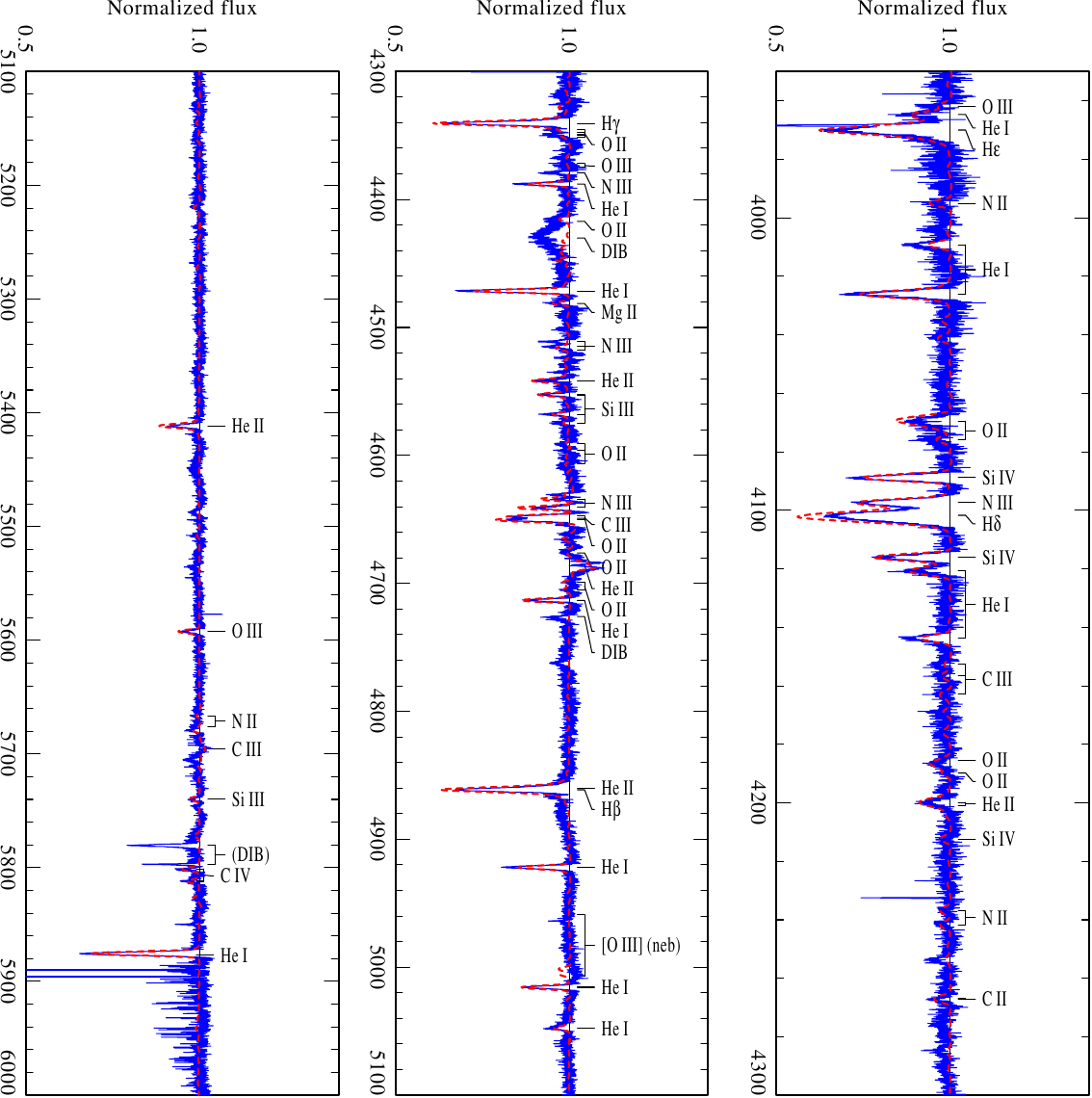}
    \caption{Optical spectra (blue) taken at orbital phase $\phi \sim 0$ and high-soft state compared to best-fit model spectra (red). }
    \label{fig:spectralfitHermes}    
\end{figure*}

\begin{figure*}
    \centering
    \includegraphics[width=1.\linewidth,angle=90]{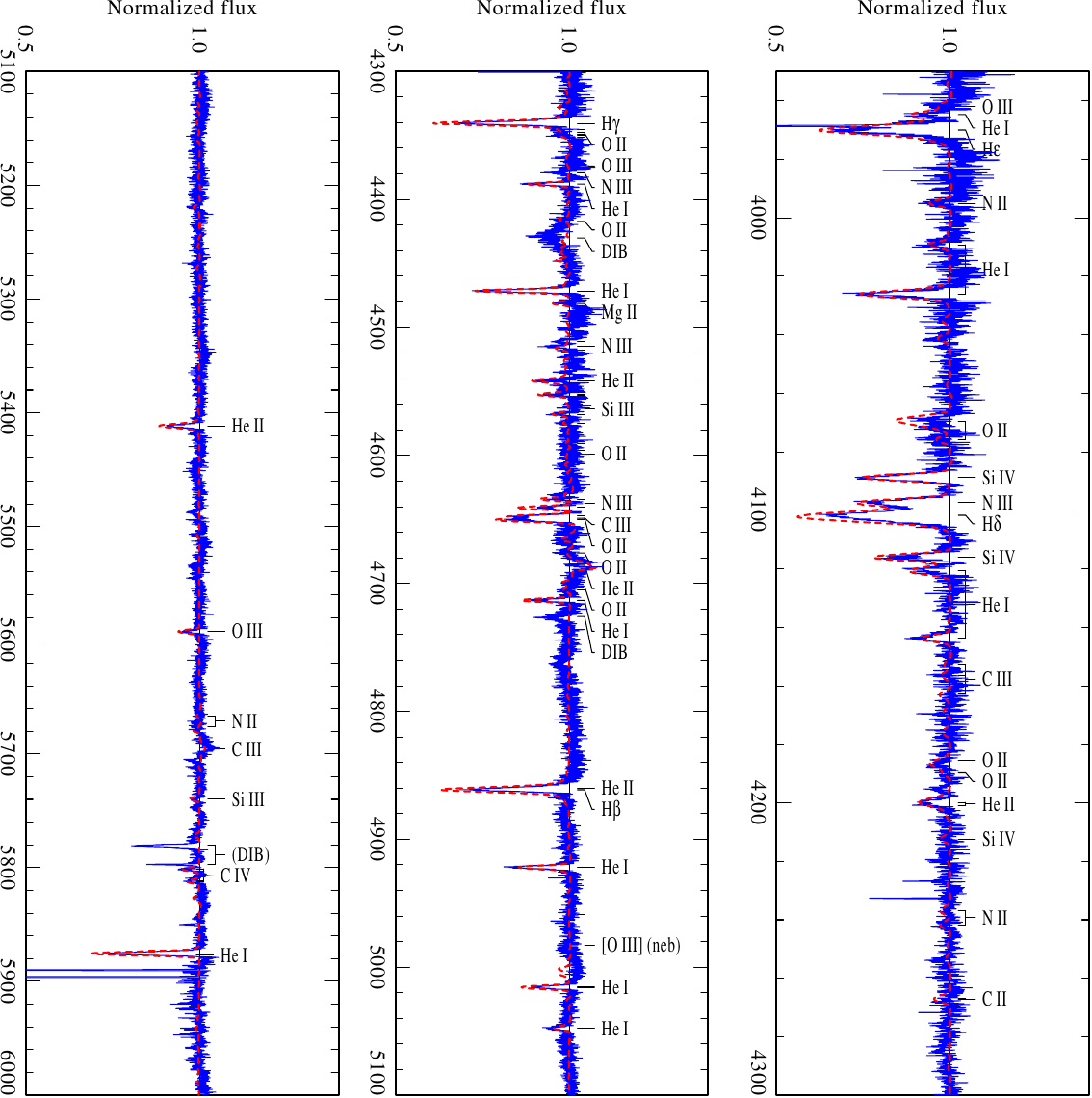}
    \caption{Optical spectra (blue) taken at orbital phase $\phi \sim 0.5$ and high-soft state compared to best-fit model spectra (red). }
    \label{fig:spectralfitHermes05}    
\end{figure*}

\begin{figure*}
    \centering
    \includegraphics[width=1.\linewidth,angle=90]{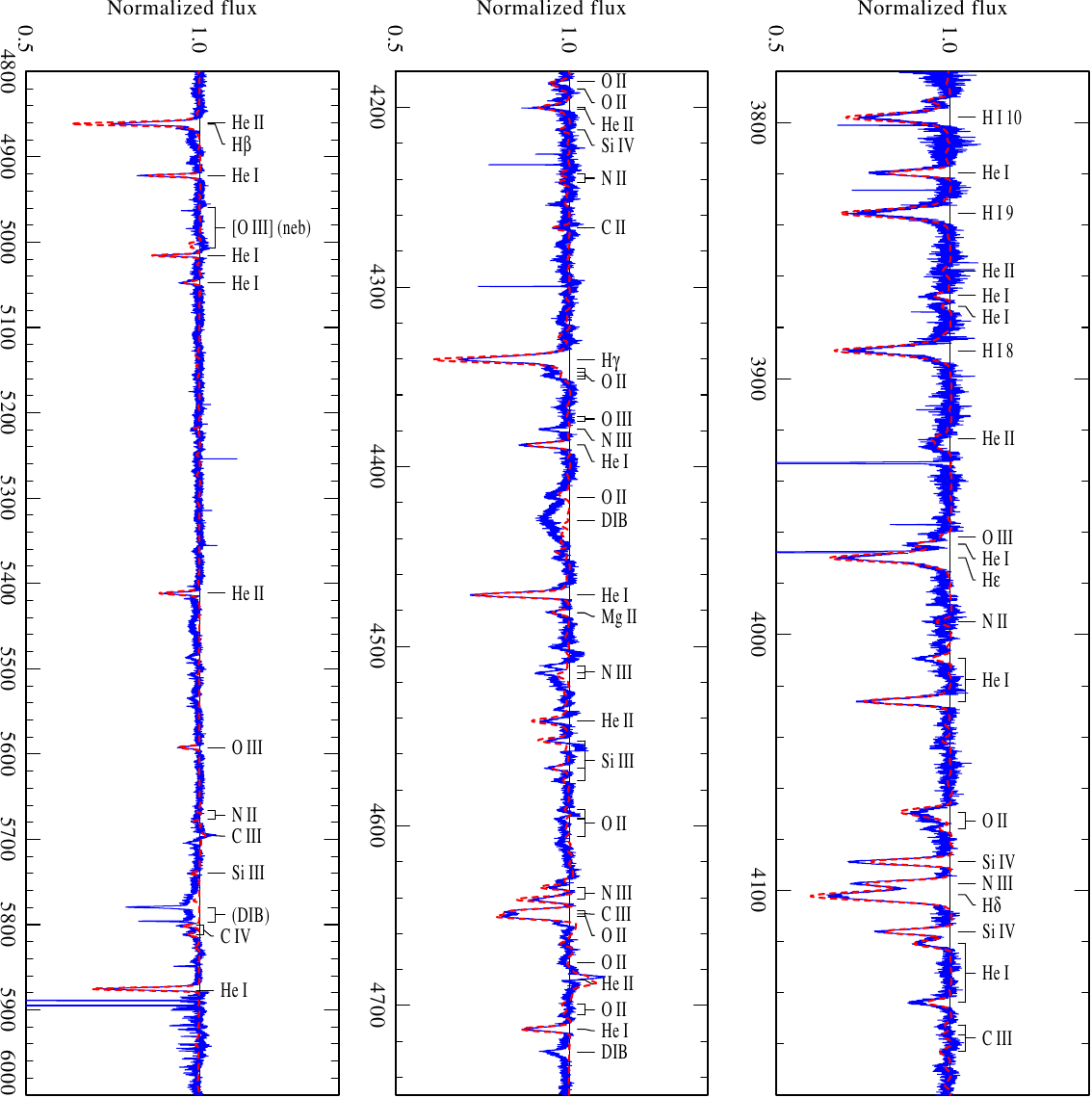}
    \caption{Optical spectra (blue) taken at orbital phase $\phi \sim 0$ and low-hard state compared to best-fit model spectra (red).}
    \label{fig:spectralfitIacob}
    
\end{figure*}

\begin{figure*}
    \centering
    \includegraphics[width=1.\linewidth,angle=90]{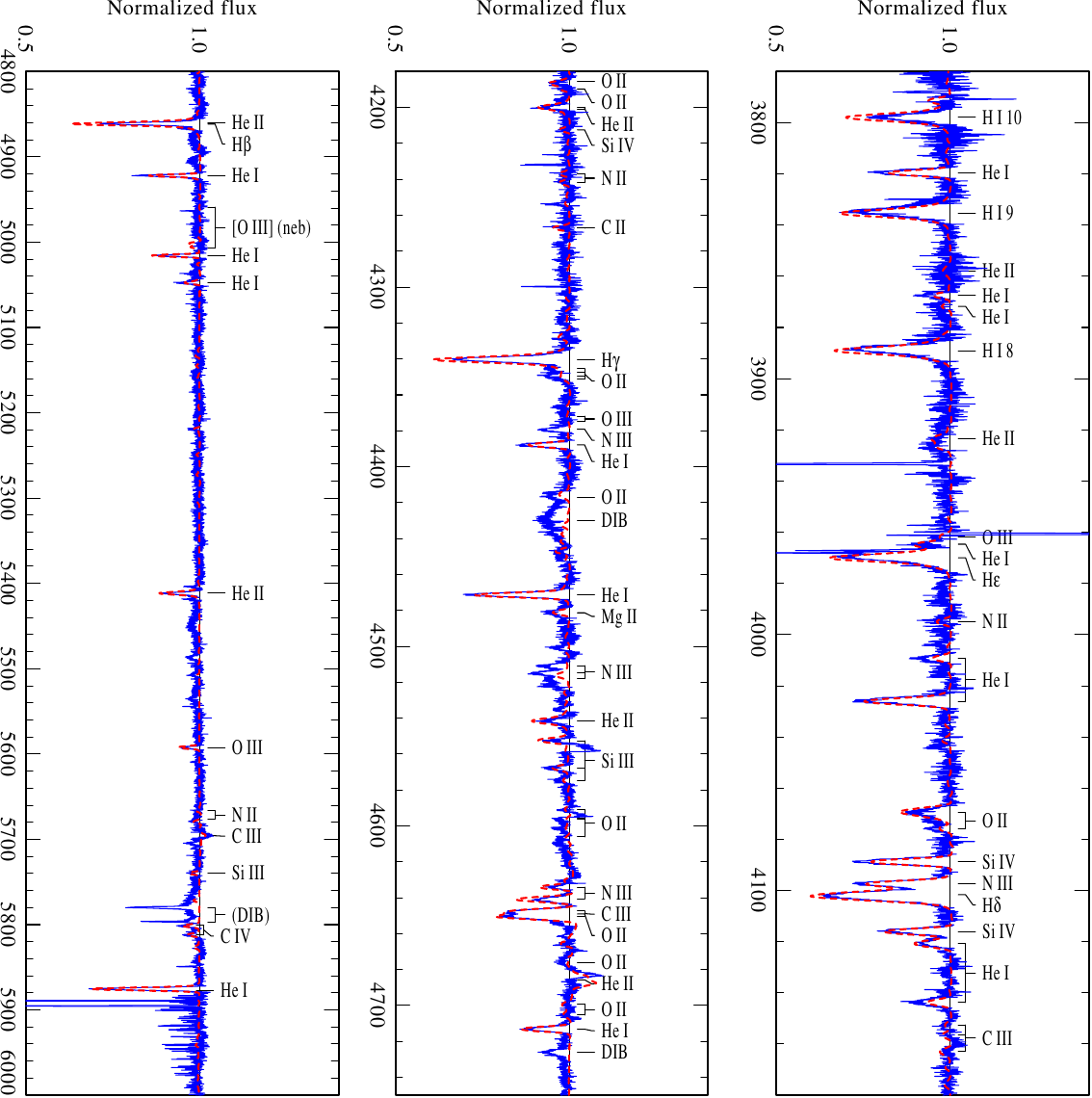}
    \caption{Optical spectra (blue) taken at orbital phase $\phi \sim 0.5$ and low-hard state compared to best-fit model spectra (red).}
    \label{fig:spectralfitIacob05}
    
\end{figure*}

\begin{figure*}
    \centering
    \includegraphics[width=0.34\linewidth,angle=90]{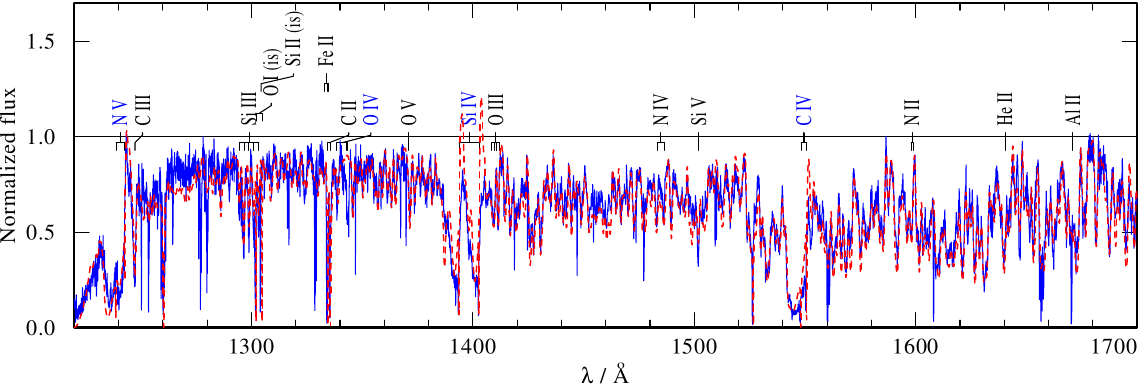}
    \includegraphics[width=0.34\linewidth,angle=90]{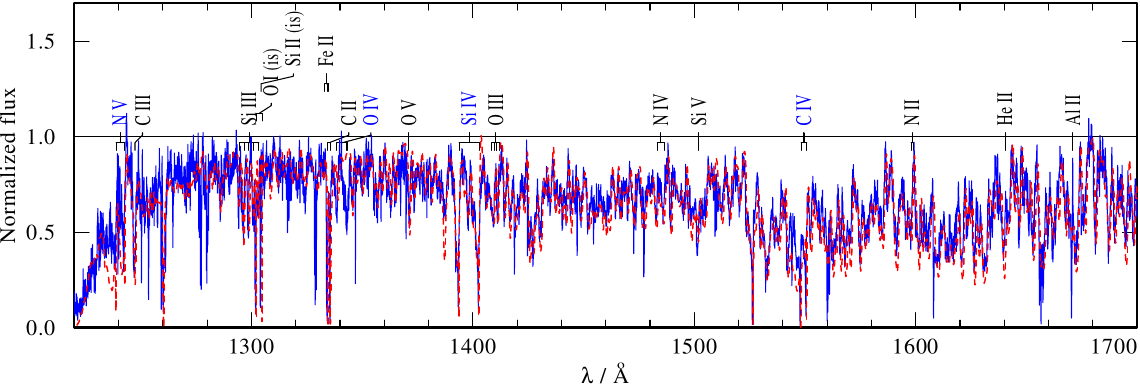}
    \caption{HST UV spectrum taken at $\phi\approx0$ (upper) and $\phi\approx0.5$ (lower) compared to best-fit model spectra.}
    \label{fig:UVfull}
\end{figure*}

\begin{figure*}
    \centering
    \includegraphics[width=1\linewidth]{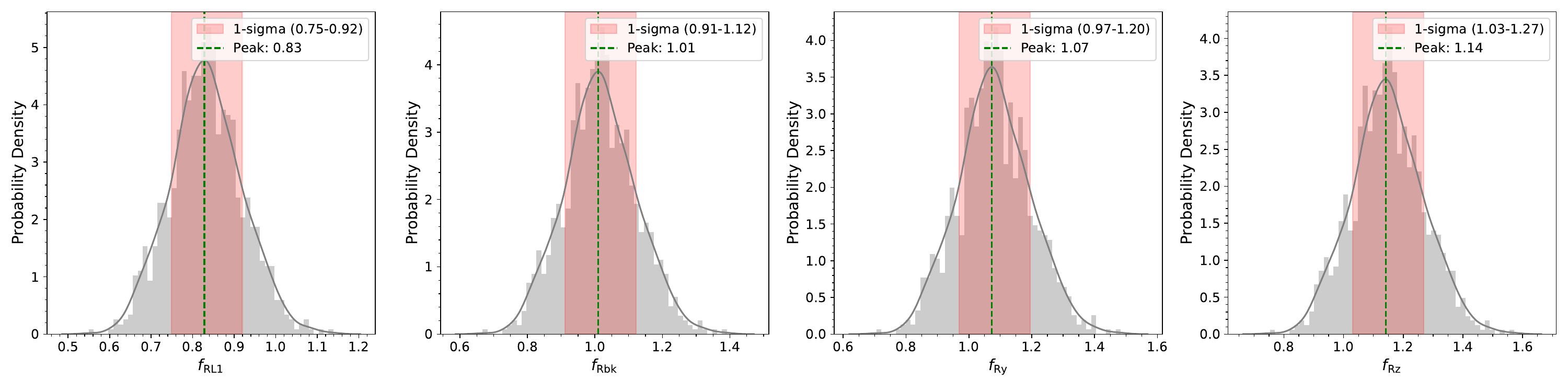}
    \caption{Probability distributions of Roche-lobe filling factors ($f_\mathrm{RL1}$, $f_\mathrm{Rbk}$, $f_\mathrm{Ry}$, $f_\mathrm{Rz}$). Roche-lobe radii in various directions were calculated using the \citet{Leahy2015} formalism, with mass ratio ($q$) distributions derived from Monte Carlo simulations incorporating the BH mass probability distribution (Fig.\, \ref{fig:MBH_distribution}). Shaded regions represent 1-sigma confidence intervals, and dashed green lines indicate distribution peaks.}
    \label{fig:fRL_hist}
\end{figure*}
 
\end{appendix}

\end{document}